\documentclass[journal]{IEEEtran}

\usepackage{cite}
\usepackage{amsmath,amssymb,amsfonts}
\usepackage{graphicx}
\usepackage{textcomp}
\usepackage{wrapfig}
\usepackage{graphicx}
\usepackage{textcomp}
\usepackage{svg}
\svgsetup{inkscapelatex=true}
\usepackage{multirow}
\usepackage{array}      
\usepackage{graphicx}
\usepackage{tabularx,booktabs}
\usepackage{algpseudocode}
\usepackage{comment}
\usepackage{xcolor}
\usepackage{hyperref}
\usepackage{cleveref}
\usepackage{ragged2e}
\usepackage{bm}

\renewcommand{\Re}{\mathop{\mathrm{Re}}\nolimits}
\renewcommand{\Im}{\mathop{\mathrm{Im}}\nolimits}

\usepackage{subcaption}

\def\BibTeX{{\rm B\kern-.05em{\sc i\kern-.025em b}\kern-.08em
    T\kern-.1667em\lower.7ex\hbox{E}\kern-.125emX}}

\definecolor{abstractbg}{rgb}{0.89804,0.94510,0.83137}
\begin{document}
\title{Gap-Induced Sidelobes in Multiband Wi-Fi CSI and Their Impact on Multipath Delay Estimation}
\author{Ana~Jeknić,~\IEEEmembership{Graduate Student Member,~IEEE,}
        Mihael~Mohorčič,~\IEEEmembership{Senior Member,~IEEE,}
        Carolina~Fortuna,
        Aleš~Švigelj,~\IEEEmembership{Senior Member,~IEEE,}
        Tomaž~Javornik,~\IEEEmembership{Senior Member,~IEEE,}
        and~Andrej~Hrovat,~\IEEEmembership{Member,~IEEE}
\thanks{\emph{Corresponding author:} Ana Jekni\'{c}, e-mail: ana.jeknic@ijs.si.}}

\maketitle

\begin{abstract}
Leveraging channel state information from multiple Wi-Fi bands has the potential to improve delay resolution for ranging and sensing when wide contiguous spectrum is unavailable. However, non-contiguous spectrum access reshapes the spectral window, producing sidelobes and secondary peaks that can mask closely spaced multipath components. Prior work either designs the bands, which fixed Wi-Fi channelization does not allow, or addresses only the first multipath component. To characterize these secondary peaks, multipath delay estimation is studied in Wi-Fi-compliant multiband settings using channel state information. For a two-path model with unknown complex gains and delays, the Cramér-Rao lower bound for delay separation is derived and analyzed, confirming the benefit of larger frequency aperture at every separation considered, while revealing pronounced, separation-dependent oscillations driven by gap geometry and coupling of the two multipath components.
Increasing aperture sharpens the mainlobe but also intensifies sidelobes and multipath interference, yielding competing peaks and a dominant peak shifted from the true delay. For the single-path matched filter, this shift is a deterministic offset governed by the slope of the autocorrelation function (ACF) of the allocation at the separation. A closed-form expression is derived for this offset, predicting it to within $21$~ps for the analyzed channel combinations. Since the spectrum allocation enters both the bound and the offset, the impact of its gaps can be assessed for any legal aggregation before a measurement is taken.
\end{abstract}


\begin{table*}[t]
\centering
\caption{Positioning of this work within the delay-estimation literature.
$\bullet$~object of study; $\circ$~enters derivations or discussion only; blank, absent.}
\label{tab:related}
\footnotesize
\setlength{\tabcolsep}{8pt}
\newcommand{\Y}{$\bullet$}
\newcommand{\I}{$\circ$}
\newcommand{\N}{}
\begin{tabular}{@{}l l l c c ccc@{}}
\toprule
 & & & \textbf{Delay} & \textbf{Multipath} & \multicolumn{3}{c}{\textbf{Bounds}} \\
\cmidrule(lr){6-8}
\textbf{Work} & \textbf{Allocation} & \textbf{Estimand} & \textbf{response} & \textbf{bias} & \textbf{CRLB} & \textbf{MCRB} & \textbf{global} \\
\midrule
\multicolumn{8}{@{}l}{\textbf{A. Multiband Sensing and Estimators (Section~\ref{sec:related_estimators})}} \\
Kazaz \emph{et al.}~\cite{kazaz2021delay}    & multiband, free       & delay vector & \N & \N & \Y & \N & \N \\
Hu \emph{et al.}~\cite{hu2025multiband}      & multiband, free       & first path, position     & \N & \N & \I & \N & \N \\
\addlinespace[3pt]
\multicolumn{8}{@{}l}{\textbf{B. Resolution Limits and Spectrum Design (Section~\ref{sec:related_bounds})}} \\
Dun \emph{et al.}~\cite{dun2021design}       & multiband, designed   & first path   & \I & \Y & \Y & \I & \N \\
Wan \emph{et al.}~\cite{wan2024fundamental}  & multiband, designed   & separation   & \I & \N & \Y & \N & \Y\rlap{$^{\dagger}$} \\
Graff, Humphreys~\cite{graff2025ziv}         & single band, designed & first path   & \Y & \N & \Y & \N & \Y \\
\addlinespace[3pt]
\multicolumn{8}{@{}l}{\textbf{C. Model Mismatch and Multipath Bias (Section~\ref{sec:related_mismatch})}} \\
Ortega \emph{et al.}~\cite{ortega2023gnss}   & single band           & first path   & \Y & \Y & \N & \Y & \N \\
Navarro \emph{et al.}~\cite{navarro2026ofdm} & single band           & first path   & \Y & \Y & \Y & \Y & \N \\
\midrule
\textbf{This work} & \textbf{multiband, fixed} & \textbf{separation} & \Y & \Y & \Y & \I & \N \\
\bottomrule
\addlinespace[3pt]
\multicolumn{8}{@{}l@{}}{\footnotesize $^{\dagger}$The global bound in~\cite{wan2024fundamental} is derived under a one-path model, with the carrier phase as nuisance.}
\end{tabular}
\end{table*}

\section{Introduction}
\label{sec:introduction}
\IEEEPARstart{W}{ireless} sensing systems built on communication-centric standards inherit the spectrum access rules, radio front-end constraints, and signal structures of those standards. High-resolution sensing benefits from large bandwidth or frequency aperture, but practical devices often operate over fragmented, non-contiguous allocations due to regulatory constraints and standard-defined channelization. Consequently, modern wireless standards are evolving towards multiband operation \cite{saeidi2023multi, 10438479, rasti2022evolution}. For example, IEEE 802.11be (Wi-Fi 7) has already introduced this as multi-link operation (MLO) that enables channel aggregation across the 2.4, 5, and 6 GHz bands~\cite{9090146}, while the forthcoming IEEE 802.11bn (Wi-Fi~8) continues multiband and MLO approach~\cite{galati2024will}.

\begin{figure}[!t]
\centering
\includegraphics[trim=10 10 5 10, clip, width=\columnwidth]{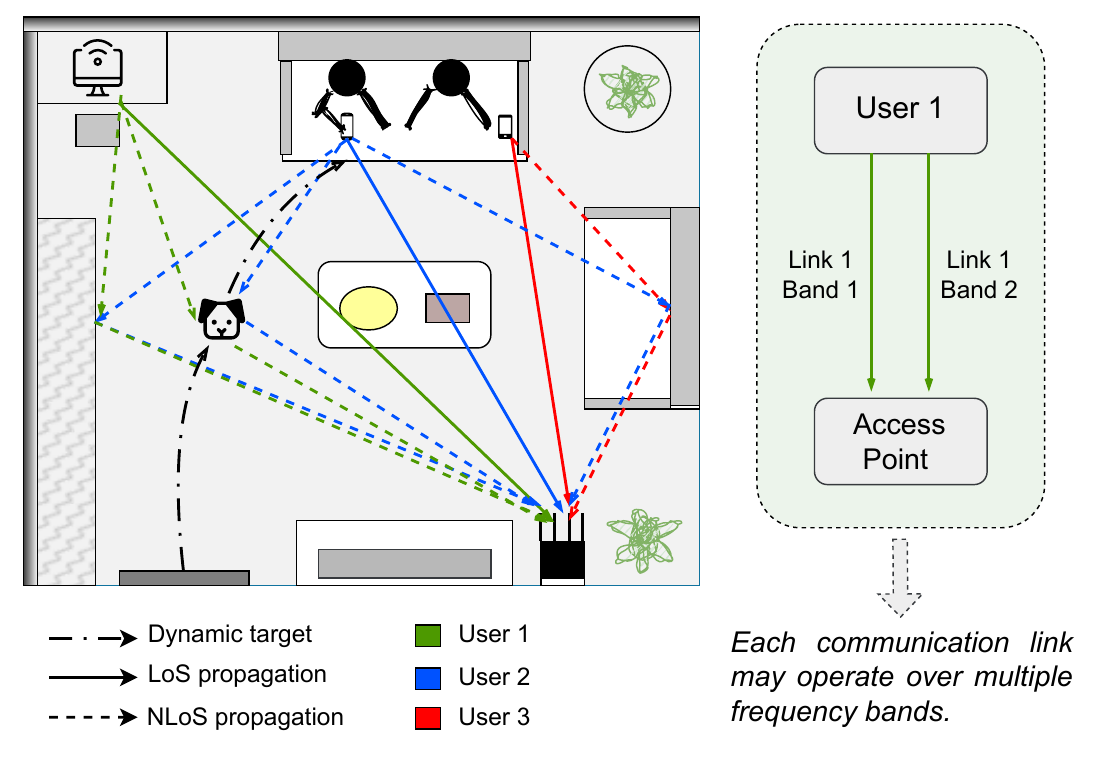}
\caption{Multiband Wi-Fi sensing scenario: each communication link between a user and the access point may operate over multiple frequency bands, while a target is observed through propagation paths.}
\label{fig:scenario}
\end{figure}

In communication-centric integrated sensing and communication (ISAC) systems, sensing estimates properties of the propagation environment from channel state information (CSI) measured over the communication link. Multiband sensing then refers to performing sensing on non-contiguous channels or bands jointly (Fig.~\ref{fig:scenario}), to synthesize a larger effective bandwidth and improve delay resolution \cite{kazaz2021delay, pegoraro2024hisac, xiao2025achieving, kato2025multi}. Since many localization and environment-characterization tasks ultimately rely on resolving the delays of multipath components (MPCs), i.e., the copies of the transmitted signal that arrive over different propagation paths, a key question is how multiband and non-contiguous spectrum impacts their estimation.

In a single band, the nominal delay resolution is defined by the occupied bandwidth, but the delay response is shaped by a mainlobe and sidelobes that decay away from it. Using two separated bands also introduces a frequency aperture that is wider than the bandwidth occupied by the individual bands. The delay response is then a combination of the two band delay responses, with a different sidelobe structure. Every MPC is observed through this combined delay response, so whether two of them can be distinguished depends on the sidelobe level that remains at the position of the adjacent mainlobes. In conventional radio-signal ranging, only the delay of the first-arriving MPC is estimated, typically the line-of-sight (LoS) component, while later MPCs are treated as interfering nuisances to be suppressed. In contrast, in environmental sensing and imaging, these MPCs also bear the primary information about the environment. Performance therefore depends not only on individual path detection, but also on whether closely spaced components can still be resolved under the gap-modified delay response, and how well.

By applying the Cramér-Rao lower bound (CRLB) approach, it is possible to mathematically determine how precisely the separation of two MPCs can be estimated under additive white Gaussian noise (AWGN). In this paper, we analyze this bound under standardized multiband IEEE 802.11 channelization, together with the autocorrelation function (ACF) that shapes the sidelobe structure of a multiband allocation. The main contributions of this work are as follows:
\begin{itemize}
\item The CRLB for the delay separation is derived for multiband allocations, with its small- and large-separation limits identified and its oscillation on the scale of the subband center spacing explained.
\item The ACF of the multiband spectral window is obtained in closed form, where the individual subbands determine its envelope and their center spacing determines the oscillation beneath it.
\item The deterministic offset of a single-path correlator applied to a two-path channel is predicted in closed form, to within $21$~ps for both delays and their separation. The prediction shows that a wider aperture narrows the peak but does not necessarily reduce the offset, as the gap also determines where each MPC falls on the sidelobes of the other.
\end{itemize}

The remainder of the paper is organized as follows. Section~\ref{sec:Related} places this work in relation to the relevant literature. Section~\ref{sec:Framework} introduces the channel and observation models, formulates the problem, and defines the multiband Wi-Fi scenarios used throughout the paper. 
Section~\ref{sec:CRLB} derives and analyzes the CRLB for a representative two-path case.
Section~\ref{sec:Sidelobes} then investigates the delay-domain sidelobe structure and matched-filter offset induced by spectral gaps, and discusses their relation to CRLB. Finally, Section~\ref{sec:conclusion} concludes the article and outlines directions for future work.

\section{Related Work}
\label{sec:Related}

Multiband operation of wireless communication systems brings two benefits to sensing and localization: diversity across separated frequencies and a frequency aperture wider than a single channel provides, which can yield finer delay resolution~\cite{armenta2024wireless, li2025sensing, ding2023multi, picazo2026802}. The delay estimation problem then falls into two categories. One treats the spectrum as a design variable and selects the best combination of bands~\cite{dun2020sparse, dun2021design} so that the delay can be estimated as precisely as possible. The other takes the spectrum as given and focuses either on the estimator~\cite{kazaz2021delay, wan2023multiband, hu2025multiband} or on the limits any estimator can achieve~\cite{wan2024fundamental}. In both cases, the quantity estimated is almost always the first path, typically the LoS component, and the subsequent MPCs are treated as a nuisance or suppressed. In this paper, the allocation is set by the standard rather than chosen, and in sensing applications, the subsequent components constitute the primary estimands of interest. We review the literature across three interrelated topics summarized in Table~\ref{tab:related}: multiband channel state information (CSI) splicing and estimator formulations, theoretical resolution bounds and spectrum design, and delay-domain sidelobes under model mismatch.

\subsection{Multiband Sensing and Estimator Formulations}
\label{sec:related_estimators}

Multiband sensing builds on
multiband splicing~\cite{dimce2024survey,dimce2025reconsidering}, which combines channel measurements collected over separated frequency allocations to emulate a wider effective bandwidth. In the delay domain this fusion targets wideband-like channel characterization, through channel impulse response (CIR) reconstruction or parametric estimation of path delays and complex gains, with time-of-arrival (ToA) estimation as a central special case for ranging and sensing. Its main practical enabler is coherence, since independent oscillators and unsynchronized chains induce phase and timing offsets~\cite{pegoraro2024hisac,wu2024sensing}, and multiband operation adds the requirement of consistent inter-band phase across allocations~\cite{dimce2024survey}.

On a given allocation, estimator families are surveyed in~\cite{dimce2024survey,wan2024ofdm} and compared for sub-7~GHz Wi-Fi in~\cite{meneghello2023wi} under commodity bandwidth constraints. Maximum-likelihood (ML) formulations~\cite{wan2023multiband,hu2025multiband} face a highly multimodal objective and search it globally from a coarse first estimate. Subspace methods instead estimate the delays from the signal subspace of the CSI~\cite{kazaz2021delay}, and sparse-recovery methods fit a few dominant delays on a grid~\cite{dimce2025reconsidering}. Crucially, these estimators treat the spectrum as unconstrained or free (representative entries in part~A of Table~\ref{tab:related}). Under non-contiguous spectrum, CIR broadening and delay-error trends emerge that follow the aperture and gap placement~\cite{jeknic2025effects}, necessitating a delay-domain analysis under fixed standard channelization.

\subsection{Resolution Limits and Spectrum Design}
\label{sec:related_bounds}

Performance limits are commonly studied through estimation-theoretic bounds, whose interpretation depends on the estimation regime assumed. The CRLB characterizes the smallest variance attainable by locally unbiased estimators through the Fisher information~\cite{kay1993fundamentals,stoica2005spectral}, whereas practical estimators are often only approximately unbiased, or deliberately biased through discretization, regularization or decision rules to reduce mean-square error (MSE)~\cite{kay1993fundamentals,lehmann1998theory}. For multiband allocations, the CRLB has been derived for the delay vector on a given allocation~\cite{kazaz2021delay} and, in closed form, for the delay separation together with the corresponding separation resolution limit (SRL)~\cite{wan2024fundamental}, where resolvability collapses periodically at multiples of the inverse subband spacing, but does not examine what shapes the oscillation. Resolution limits are surveyed in~\cite{wan2024ofdm}. Apart from~\cite{wan2024fundamental}, the quantity bounded is the first-arriving MPC, typically the LoS component, with the remaining components entering as nuisance parameters or, as in~\cite{kazaz2021delay}, assumed resolved so that the bias can be ignored.

Being local, the CRLB does not describe the large errors that follow when an ambiguous response leads to the wrong peak, and global bounds of the Ziv-Zakai family are used to reach that regime~\cite{wan2024fundamental,graff2025ziv}. In both, however, the bound is obtained under a one-path model, so the second path is absent rather than a nuisance parameter, and extending it beyond one path is reported as computationally prohibitive~\cite{wan2024fundamental}. On the design side, candidate subbands or pilot patterns are selected by optimizing CRLB criteria under LoS-plus-reflection models~\cite{dun2020sparse,dun2021design} or the Ziv-Zakai bound~\cite{graff2025ziv}. Part~B of Table~\ref{tab:related} summarizes these works, which ultimately aim to optimize the allocation. While effective when spectrum can be freely engineered, this approach does not apply to standard-compliant IEEE~802.11 allocations, where channelization is rigid.

\subsection{Delay-Domain Response, Model Mismatch, and Multipath Bias}
\label{sec:related_mismatch}

In Wi-Fi and other orthogonal frequency-division multiplexing (OFDM) systems, the delay domain is obtained by applying an inverse discrete Fourier transform (IDFT) to the measured channel frequency response (CFR). This yields the CIR as seen through the occupied spectrum, with each MPC carrying the corresponding sidelobe pattern. That pattern follows directly from the autocorrelation function (ACF) of the occupied spectrum~\cite{graff2025ziv}, which, in estimation terms, corresponds to the widely used correlation of the measurement with a single-path signature~\cite{navarro2026ofdm}. The same object also appears under the name point-spread function~\cite{kou2026dualband}. We refer to it as the ACF, and use delay response when the emphasis is on its shape in delay.

A critical source of error in practical systems is model mismatch. When an under-parameterized single-path matched-filter model is applied to a multipath channel, the estimator converges to a pseudo-true parameter instead of the delay. Its variance is bounded by the misspecified Cramér-Rao bound (MCRB)~\cite{richmond2015parameter,fortunati2017performance}, and its error decomposes as $\mathrm{MSE}=\mathrm{variance}+\mathrm{bias}^{2}$, where the deterministic bias cannot be captured by correctly specified variance bounds. This phenomenon has been examined for global navigation satellite system (GNSS) synchronization under interference~\cite{ortega2023gnss} and contiguous-OFDM ranging~\cite{navarro2026ofdm} (bottom rows of Table~\ref{tab:related}), where the bias follows the slope of the autocorrelation at the path separation. For contiguous signals, that response has one mainlobe followed by decaying sidelobes.  Under gapped spectrum, however, the sidelobes become more pronounced, so that the offset depends on where each MPC falls on the sidelobes of the other.

Table~\ref{tab:related} summarizes how this work integrates these axes. While prior studies focus on free or engineered spectrum, single-path bounds, or contiguous mismatch, this paper analyzes fixed, standard-compliant IEEE~802.11 multiband allocations, establishing a closed-form CRLB on delay separation alongside explicit characterization of gap-induced sidelobes and deterministic matched-filter offset.

\begin{table}[t]
  \centering
  \caption{Summary of Principal Notation}
  \label{tab:notations}
  \setlength{\tabcolsep}{4pt}
  \renewcommand{\arraystretch}{1.05}
  \footnotesize
  \begin{tabularx}{\columnwidth}{@{}l>{\RaggedRight\arraybackslash}X@{}}
    \toprule
    \textbf{Symbol} & \textbf{Description} \\
    \midrule
    \multicolumn{2}{@{}l}{\textbf{Grid, Allocation, and Channel Model (Section~\ref{sec:Framework})}} \\
    $s$ & Scenario index, $s\in\{\mathrm{A1}-\mathrm{A3},\mathrm{B1}-\mathrm{B3}\}$ (Table~\ref{tab:scenarios}) \\
    $f[n],\ \Delta f$ & Frequency of grid point $n$; spacing, $78.125$~kHz \\
    $N_f,\ B_{\mathrm{tot}}$ & Number of grid points; aperture $(N_f-1)\Delta f$ \\
    $a_s[n]$ & Spectral mask; $\{0,1\}$-valued when rectangular \\
    $\mathcal{K}_s,\ N_s$ & Occupied-subcarrier set and its size, $N_s=|\mathcal{K}_s|$ \\
    $B,\ B_{\mathrm{sb}}$ & Total occupied and per-subband bandwidth \\
    $f_{c,i},\ \Delta f_c$ & Center of subband $i$; spacing $f_{c,2}-f_{c,1}$ \\
    $L,\ \alpha_\ell,\ \tau_\ell$ & Number of paths; complex gain and delay of path $\ell$ \\
    $\Delta\tau$ & Delay separation $\tau_2-\tau_1>0$ \\
    $\phi$ & Relative phase of the two MPCs, $\arg(\alpha_2/\alpha_1)$ \\
    $Y_s[n],\ W_s[n]$ & Measured CFR on subcarrier $n$ and its additive noise \\
    $\sigma_s^2$ & Per-subcarrier noise variance over $\mathcal{K}_s$ \\
    \midrule
    \multicolumn{2}{@{}l}{\textbf{Fisher Information and Bounds (Section~\ref{sec:CRLB})}} \\
    $\boldsymbol{\theta}$ & Parameters $[\tau_1\ \tau_2\ \alpha_{1\mathrm{Re}}\ \alpha_{1\mathrm{Im}}\ \alpha_{2\mathrm{Re}}\ \alpha_{2\mathrm{Im}}]^{T}$ \\
    $\mathbf{I}_s(\boldsymbol{\theta}),\ I_{ij}$ & $6\times6$ FIM and its entries \\
    $\mathbf{I}_{\tau\tau},\mathbf{I}_{\tau\alpha},\mathbf{I}_{\alpha\alpha}$ & Delay, delay-gain, and gain blocks of the FIM \\
    $\mathbf{I}_{\mathrm{eff}}(\boldsymbol{\tau}),\ \tilde{I}_{ij}$ & Effective delay information and its entries \\
    $\mathrm{CRLB}(\Delta\tau)$ & CRLB on the delay separation \\
    $\mathcal{V}_s$ & Weighted spread of $f[n]^{2}$ in the small-$\Delta\tau$ bound \\
    $\kappa(\alpha_1,\alpha_2)$ & Gain factor of the small-$\Delta\tau$ bound \\
    \midrule
    \multicolumn{2}{@{}l}{\textbf{Delay Response and Matched-Filter Offset (Section~\ref{sec:Sidelobes})}} \\
    $g_s(u)$ & Single-path delay response (ACF) at offset $u$ \\
    $G_i(u)$ & Basebanded response of subband $i\in\{1,2\}$ \\
    $B_{\mathrm{rms}}$ & RMS spread of occupied subcarriers about their centroid \\
    $T_s(\tau)$ & Noise-free single-path scan of the two-path CFR \\
    $\hat\tau_\ell,\ b_\ell$ & Located peak and its offset $\hat\tau_\ell-\tau_\ell$ \\
    $\widehat{\Delta\tau}_{\mathrm{MF}},\ b_\Delta$ & Scan separation $\hat\tau_2-\hat\tau_1$ and its offset $b_2-b_1$ \\
    $\hat b^{(1)}_\ell$ & First-order prediction of $b_\ell$ \\
    $\hat b_\ell,\ \hat b_\Delta$ & Second-order predictions of $b_\ell$ and $b_\Delta$ \\
    $\lambda_s(u)$ & Normalized slope $|g_s'(u)|/\big(2\pi g_s(0)B_{\mathrm{rms}}\big)$ \\
    $\bar g_s,\ \chi_s$ & Level $g_s(\Delta\tau)/g_s(0)$ and curvature $g_s''(\Delta\tau)/g_s''(0)$ of the ACF \\
    $\gamma_s(\Delta\tau)$ & Normalized sidelobe level $|\bar g_s|=|g_s(\Delta\tau)|/|g_s(0)|$ \\
    $\eta_s$ & Share of delay information kept, $\tilde{I}_{\ell\ell}/I_{\ell\ell}$ \\
    $\varrho_s$ & Normalized coupling $\tilde{I}_{12}/\sqrt{\tilde{I}_{11}\tilde{I}_{22}}$ \\
    $r_\ell$ & Gain ratio $|\alpha_{\bar\ell}/\alpha_\ell|$, with $\bar\ell$ the other MPC \\
    $v_\ell$ & Single-path bound on $\tau_\ell$ \\
    \bottomrule
  \end{tabularx}
\end{table}

\section{Analysis Framework}
\label{sec:Framework}

This section introduces the signal and observation models used throughout the paper. A unified frequency-domain formulation is adopted to describe both single-band and multiband Wi-Fi sensing scenarios on a common global frequency grid, while explicitly accounting for standard-compliant channelization and non-contiguous spectrum occupancy. This framework provides a consistent basis for the subsequent analysis of delay estimation performance, including both estimation-theoretic bounds and delay-domain behavior under gapped spectrum.

The symbols used across sections are summarized in Table~\ref{tab:notations}, grouped by the section that defines them.

\subsection{Channel Model and Global Frequency Grid}
\label{subsection:model}

We model CIR as a sum of $L$ propagation paths:
\begin{equation}
h(t) = \sum_{\ell=1}^{L} \alpha_{\ell}\,\delta(t-\tau_{\ell}),
\label{eq:cir_model}
\end{equation}
where $\alpha_{\ell}\in\mathbb{C}$ is the complex gain and $\tau_{\ell}\in\mathbb{R}_{+}$ is the delay of path $\ell$. The corresponding CFR is
\begin{equation}
H(f) = \sum_{\ell=1}^{L} \alpha_{\ell}\,e^{-j2\pi f \tau_{\ell}}.
\label{eq:cfr_model}
\end{equation}

\smallskip
To describe both single-band and multiband measurements in a unified way, we construct a global frequency grid spanning the overall measurement aperture $[f_{\mathrm{start}},\,f_{\mathrm{stop}}]$:
\begin{equation}
f[n] = f_{\mathrm{start}} + n\Delta f,\qquad n=0,1,\dots,N_f-1,
\label{eq:global_grid}
\end{equation}
where $f_{\mathrm{start}}$ and $f_{\mathrm{stop}}$ denote the grid endpoints, $\Delta f$ is the subcarrier frequency spacing, and $N_f$ is the total number of subcarriers on the grid. The corresponding total aperture is:
\begin{equation}
B_{\mathrm{tot}} = f_{\mathrm{stop}}-f_{\mathrm{start}} = (N_f-1)\Delta f.
\label{eq:aperture}
\end{equation}
While the framework can be extended to accommodate band-dependent spacings, we adopt a common $\Delta f$ across all bands.

\subsection{Multiband Spectral Mask and CSI Observations}
\label{subsec:observations}

A multiband scenario $s$ activates one or more Wi-Fi channels within this global grid. Spectral occupancy is described through a scenario-dependent spectral mask $a_s[n]$ defined on the full grid, $n=0,\dots,N_f-1$. In a standard-compliant allocation the mask jointly captures the activated channel(s), the omitted subcarriers within each channel bandwidth, the guard and DC-null subcarriers (i.e., empty centre subcarriers to avoid direct current offset), and the Wi-Fi spectral shaping~\cite{10979691}, which is approximately flat in-band and decays in the band-edge roll-off. For subcarriers that are not used or are suppressed, $a_s[n]=0$.

In the multiband case, the spectral mask can be decomposed into contributions from the individual channels. For a two-channel aggregation, we write:
\begin{equation}
a_s[n] = a_{s,1}[n] + a_{s,2}[n],
\label{eq:mask_decomposition}
\end{equation}
where $a_{s,1}[n]$ and $a_{s,2}[n]$ denote the spectral masks of two separate Wi-Fi channels. Each sub-mask is supported on its associated channel, centred at $f_{c,i}$ with $f_{c,1}<f_{c,2}$, and is zero outside it. The spectral gap between the channels is therefore represented implicitly by the region where both $a_{s,1}[n]$ and $a_{s,2}[n]$ are equal to zero.

The following analysis adopts a rectangular mask instead of the standard-compliant one, for analytical tractability and so that the delay-domain behavior remains attributable to the occupied channels and the gap between them. Every subcarrier of an activated channel is assigned $a_s[n] = 1$ and every other subcarrier $0$. This differs from the standard-compliant mask in two respects. First, the band-edge roll-off is absent, which has little effect because it decays across subcarriers that the subcarrier plan has already nulled as guards. Second, the guard and DC-null subcarriers, which a standard-compliant allocation sets to zero, are treated here as occupied. The resulting differences are quantified, and the standard-compliant case is discussed alongside the related analyses.

For notational convenience in subsequent derivations, we define the set of used subcarriers as:
\begin{equation}
\mathcal K_s = \{\, n \in \{0,\dots,N_f-1\} : a_s[n] \neq 0 \,\}.
\label{eq:Ks_definition}
\end{equation}
The set has $N_s=|\mathcal{K}_s|$ elements, and its part in channel $i$ is $\mathcal{K}_{s,i}$. Since $a_s[n]\in\{0,1\}$ under the rectangular mask, $|a_s[n]|^{2}=a_s[n]$ and $a_s$ is the indicator function of $\mathcal{K}_s$; the mask-power weights appearing in the Fisher information of Section~\ref{sec:CRLB} and in the delay response of Section~\ref{sec:Sidelobes} therefore reduce to uniform weighting over the occupied subcarriers.

Throughout, frequencies are referred to the centre of the occupied set, so that $\sum_{n\in\mathcal{K}_s}f[n]=0$. This fixes the origin of the frequency axis and involves no loss of generality: displacing the origin by $f_0$ multiplies the signature $a_s[n]e^{-j2\pi f[n]\tau_\ell}$ of the MPC at $\tau_\ell$ by the constant $e^{j2\pi f_0\tau_\ell}$, which the unknown complex gain absorbs, leaving the delay information and the delay-domain magnitudes unchanged. For the allocations in Table~\ref{tab:scenarios}, the two subbands carry equal bandwidth $B_{\mathrm{sb}}$, and the rectangular mask weights their subcarriers equally, so their centres lie symmetrically about the origin, at $\mp\Delta f_c/2$ with:
\begin{equation}
\Delta f_c=f_{c,2}-f_{c,1}.
\label{eq:dfc}
\end{equation}
The symmetric case is carried throughout; an asymmetric allocation is shown where it differs, to illustrate the generalization. 

In OFDM-based Wi-Fi, CSI is obtained from the long training field (LTF) by dividing the symbol received on each subcarrier by the known LTF symbol. On an occupied subcarrier $n\in\mathcal{K}_s$ the resulting CFR measurement is:

\begin{equation}
Y_s[n] = a_s[n]\,H\!\big(f[n]\big) + W_s[n],\qquad n\in\mathcal{K}_s.
\label{eq:cfr_fullgrid}
\end{equation}
The term  $W_s[n]\sim\mathcal{CN}(0,\sigma_s^2)$ is additive complex Gaussian noise, with the noise variance $\sigma_s^2$, uncorrelated across subcarriers and of equal variance, as follows from the per subcarrier LTF division. The noise is restricted  to the measured subcarriers by the condition $n\in\mathcal{K}_s$, and not by the spectral mask $a_s[n]$, because the mask specifies transmit spectral shaping, and does not weight noise generated in the receiver. Across the widely separated bands of a multiband allocation, the noise variance could differ, and the framework could be extended to accommodate a per-band variance. However, throughout this study $\sigma_s^2$ is kept the same for all bands, so that differences between scenarios reflect only the allocation geometry.

\begin{figure*}[!t]
  \centering
  \includegraphics[width=\textwidth]{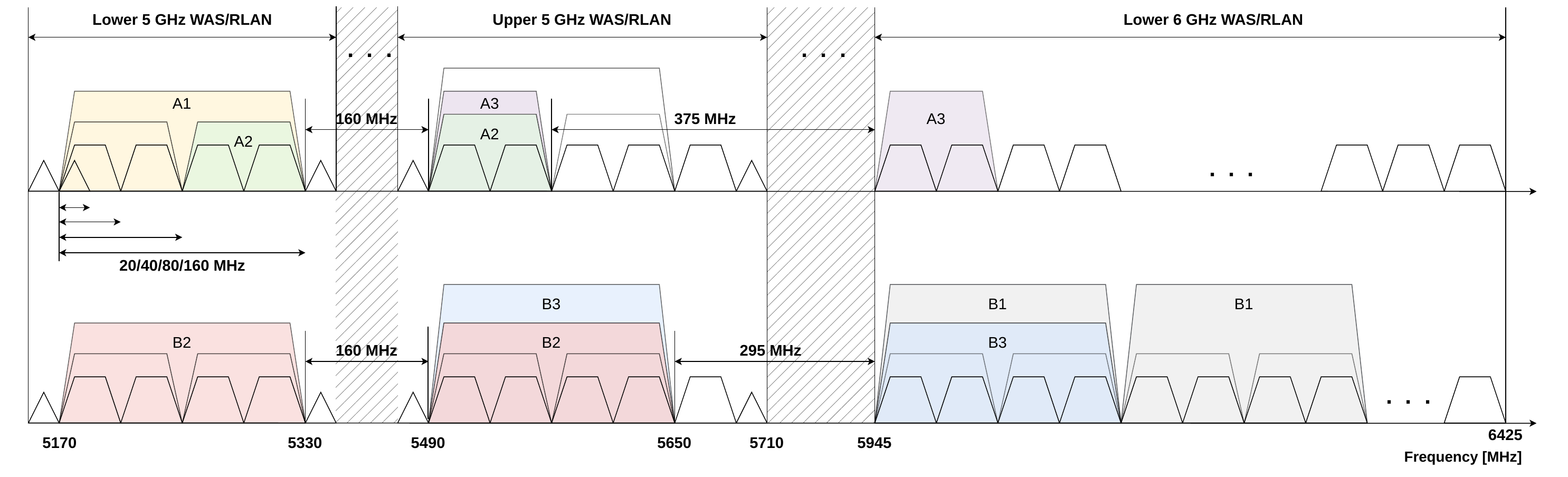}
  \caption{Simulation scenarios A1-A3 and B1-B3 positioned within the European Wi-Fi channelization framework under European Telecommunications Standards Institute (ETSI) regulations.}
  \label{fig:multiband}
\end{figure*}

The CFR measurements can be embedded into the frequency-domain vector:
\begin{equation}
{\mathbf{Y}}_s =
\begin{bmatrix}
Y_s[0] & Y_s[1] & \cdots & Y_s[N_f-1]
\end{bmatrix}^{T}.
\label{eq:embedded_vector_observations}
\end{equation}
Since no measurement exists outside $\mathcal{K}_s$, the entries in the spectral gaps between the occupied bands are set to $Y_s[n]=0$, so that neither signal nor noise is present there.

If needed, an $N_f$-point inverse DFT can be computed to obtain a discrete CIR estimate:
\begin{equation}
{y}_s[p] = \frac{1}{N_f}\sum_{n=0}^{N_f-1} {Y}_s[n]\,
e^{j2\pi np/N_f},\qquad p=0,1,\dots,N_f-1.
\label{eq:ifft_cir}
\end{equation}
This induces a discrete delay axis $t[p]=p/(N_f\Delta f)$ for $p=0,\dots,N_f-1$. Because the inverse DFT is periodic, the resulting delay-domain representation spans an unambiguous window of length $1/\Delta f$.

Given noisy multiband CFR/CSI measurements collected under scenario $s$, the goal of multipath delay estimation is to estimate the delays of the dominant MPCs, i.e., the parameter vector $\boldsymbol{\tau}=[\tau_1,\dots,\tau_L]^T$, and, depending on the estimator, also the corresponding complex gains $\boldsymbol{\alpha}=[\alpha_1,\dots,\alpha_L]^T$. Estimates can be obtained by processing the frequency-domain observations directly, or indirectly by transforming the measurements to the delay domain via Eq.~\eqref{eq:ifft_cir} and identifying the dominant components on the resulting delay axis.

\subsection{Two-Path Channel and Delay Separation Formulation}
\label{subsec:twopath_formulation}

To study how multiband spectral occupancy affects the ability to distinguish closely spaced MPCs, we set $L=2$ in Eq.~\eqref{eq:cfr_model}, which gives the two-path model:
\begin{equation}
H(f) = \alpha_1 e^{-j2\pi f\tau_1} + \alpha_2 e^{-j2\pi f\tau_2},
\label{eq:two_path}
\end{equation}
and parameterize path proximity by the delay separation:
\begin{equation}
\Delta\tau = \tau_2-\tau_1.
\label{eq:delta_tau_def}
\end{equation}
In the rest of the paper, the two paths are indexed so that $\tau_2>\tau_1$ and hence $\Delta\tau>0$. Although richer multipath models with $L>2$ are possible, the two-path setting provides a transparent baseline for interpreting resolvability effects and is therefore used throughout the subsequent analysis.

To avoid an arbitrary overall scaling, we normalize the dominant path gain to $\alpha_1=1$. We interpret $\alpha_1$ as the dominant component, typically the LoS component, otherwise the strongest one, and $\alpha_2$ as a single specular MPC (e.g., a reflection). 
For the second path, we use a representative complex gain $\alpha_2=|\alpha_2|e^{j\phi}$, with $\phi$ the phase of the second MPC relative to the first, and set $|\alpha_2|=0.7$ and $\phi=\pi/3$, i.e., $\alpha_2\approx 0.35+0.606j$ (about $-3$~dB in power relative to $\alpha_1$).
This choice without loss of generality fixes a nominal relative strength and phase so that comparisons across spectrum configurations and delay separations are evaluated under the same channel state (fixed path-gain ratio and relative phase). Alternative choices of $(|\alpha_2|,\phi)$ generally change the numerical values but do not affect the underlying formulation.

\subsection{Multiband Wi-Fi Scenarios}

Multiband sensing scenarios are constructed by activating one or two Wi-Fi channels within single or multiple bands. Each scenario $s$ is then fully specified by the occupied set $\mathcal{K}_s$, since under the rectangular mask $a_s[n]$ is its indicator. To quantify the impact of total used bandwidth, frequency aperture, and spectral gaps on multipath delay resolution, we consider six representative configurations summarized in Table~\ref{tab:scenarios} and used consistently in the following analysis. These scenarios are also shown in Fig.~\ref{fig:multiband}, which positions them within the European Wi-Fi channelization framework under ETSI regulations for the 5 \,GHz and lower-6 \,GHz bands. 

\begin{table}[t]
  \centering
  \caption{Multiband simulation scenarios. Group~A uses 160\,MHz and Group B 320\,MHz of channel bandwidth.}
  \label{tab:scenarios}
  \setlength{\tabcolsep}{3pt}
  \renewcommand{\arraystretch}{1.1}
  \footnotesize
  \begin{tabular}{@{}cccc@{}}
    \toprule
    ID & Occupied bands [GHz] & Aperture [MHz] & Gap [MHz] \\
    \midrule
    A1 & $[5.17,5.33]$                         & 160 & 0   \\
    A2$^{\ast}$ & $[5.25,5.33],\,[5.49,5.57]$           & 320 & 160 \\
    A3$^{\ast}$ & $[5.49,5.57],\,[5.945,6.025]$         & 535 & 375 \\
    \midrule
    B1 & $[5.945,6.105],\,[6.105,6.265]$       & 320 & 0   \\
    B2$^{\ast}$ & $[5.17,5.33],\,[5.49,5.65]$           & 480 & 160 \\
    B3$^{\ast}$ & $[5.49,5.65],\,[5.945,6.105]$         & 615 & 295 \\
    \bottomrule
  \end{tabular}

  \vspace{3pt}
  \begin{minipage}{\columnwidth}
  \footnotesize
  \justifying
  \noindent\emph{Note:} Scenarios marked with \(^\ast\) also include a contiguous reference allocation spanning the same aperture without the gap:
    A2\(^\ast\)\(:[5.25,5.57]\), A3\(^\ast\)\(:[5.49,6.025]\), B2\(^\ast\)\(:[5.17,5.65]\), and B3\(^\ast\)\(:[5.49,6.105]\).
  \end{minipage}
\end{table}

Scenarios in Group~A use a total occupied bandwidth of $B=160$\,MHz, while Group~B uses $B=320$\,MHz. Within each group, the same total bandwidth is placed either contiguously within one band or split across separated 5/6\,GHz allocations, producing increasingly gapped spectrum. Scenarios marked with $^{\ast}$ additionally include a contiguous reference allocation spanning the same aperture but without the gap, isolating the effect of spectral fragmentation.

\section{Cramér-Rao lower bound}
\label{sec:CRLB}

A CRLB benchmark is provided for the multiband observations in Eq.~\eqref{eq:embedded_vector_observations}. The benchmark quantifies the best possible precision of locally unbiased delay estimators under the assumed additive white Gaussian noise. The evaluation uses the two path channel model in Eq.~\eqref{eq:two_path}, with the delay separation \(\Delta\tau\) defined in Eq.~\eqref{eq:delta_tau_def}.

\subsection{CRLB Derivation}

Restricting the global-grid model in Eq.~\eqref{eq:cfr_fullgrid} to $\mathcal{K}_s$ yields the stacked observation:

\begin{equation}
  \mathbf{y}_s = \big[\,Y_s[n]\,\big]_{n\in\mathcal{K}_s}
  \in\mathbb{C}^{N_s}.
  \label{eq:crlb_y_stack}
\end{equation}
From Eq.~\eqref{eq:cfr_fullgrid}, $Y_s[n]$ includes the additive noise term $W_s[n]$. For the parameter vector $\boldsymbol{\theta}$ of Eq.~\eqref{eq:crlb_theta}, the observation satisfies:
\begin{equation}
  \mathbf{y}_s \,|\, \boldsymbol{\theta}
  \sim
  \mathcal{CN}\!\big(\boldsymbol{\mu}_s(\boldsymbol{\theta}),\,\sigma_s^2\mathbf{I}_{N_s}\big),
  \label{eq:crlb_gaussian}
\end{equation}
where $\mathbf{I}_{N_s}$ denotes the $N_s\times N_s$ identity matrix and $\sigma_s^2$ is the noise variance of Eq.~\eqref{eq:cfr_fullgrid}.

The corresponding mean can be written as:

\begin{equation}
  \boldsymbol{\mu}_s(\boldsymbol{\theta})
  = \mathbf{A}_s\,\mathbf{X}_s(\boldsymbol{\tau})\,\boldsymbol{\alpha},
  \label{eq:crlb_mu}
\end{equation}

where:
\begin{equation}
\begin{aligned}
\mathbf{A}_s & = \mathrm{diag}\!\big([a_s[n]]_{n\in\mathcal{K}_s}\big),\\
\boldsymbol{\tau} & = [\tau_1\ \tau_2]^T,\\
\boldsymbol{\alpha} & = [\alpha_1\ \alpha_2]^T,\\
\mathbf{X}_s(\boldsymbol{\tau}) & = \big[\mathbf{x}_s(\tau_1)\ \mathbf{x}_s(\tau_2)\big],\\
\mathbf{x}_s(\tau_\ell) & = \big[e^{-j2\pi f[n]\tau_\ell}\big]_{n\in\mathcal{K}_s},\quad \ell\in\{1,2\}.
\end{aligned}
\end{equation}

The parameter vector is:
\begin{equation}
  \boldsymbol{\theta}
  =
  \begin{bmatrix}
    \tau_1 & \tau_2 & \alpha_{1\mathrm{Re}} & \alpha_{1\mathrm{Im}} & \alpha_{2\mathrm{Re}} & \alpha_{2\mathrm{Im}}
  \end{bmatrix}^{T},
  \label{eq:crlb_theta}
\end{equation}
where \(\alpha_\ell=\alpha_{\ell\mathrm{Re}}+j\alpha_{\ell\mathrm{Im}}\) for \(\ell\in\{1,2\}\).
For each occupied subcarrier $n\in\mathcal{K}_s$
the corresponding entry of $\boldsymbol{\mu}_s(\boldsymbol{\theta})$ is:
\begin{equation}
  \mu_s[n]
  =
    a_s[n]\sum_{\ell=1}^{2}\alpha_\ell e^{-j2\pi f[n]\tau_\ell},
  \qquad n\in\mathcal{K}_s.
  \label{eq:crlb_mu_entry}
\end{equation}

For the model in Eq.~\eqref{eq:crlb_gaussian}, the Fisher information matrix (FIM) entries follow the standard complex Gaussian form \cite{kay1993fundamentals, dun2020sparse}:
\begin{equation}
  \big[\mathbf{I}_s(\boldsymbol{\theta})\big]_{ij}
  =
  \frac{2}{\sigma_s^2}
  \Re\!\left\{
    \frac{\partial \boldsymbol{\mu}_s^{H}}{\partial \theta_i}
    \frac{\partial \boldsymbol{\mu}_s}{\partial \theta_j}
  \right\},
  \qquad i,j\in\{1,2,3,4,5,6\}.
  \label{eq:crlb_fim_general}
\end{equation}
Defining \(\mathbf{d}_i = \partial \boldsymbol{\mu}_s / \partial \theta_i\), Eq.~\eqref{eq:crlb_fim_general} can be written equivalently as \(I_{ij}=\big[\mathbf{I}_s(\boldsymbol{\theta})\big]_{ij}=\frac{2}{\sigma_s^2}\Re\{\mathbf{d}_i^{H}\mathbf{d}_j\}\). 
Derivative vectors and closed form FIM entries are listed in Appendix~\ref{app:fim_entries}. With the frequency centring explained in Section~\ref{subsec:observations}, the linear frequency sums vanish, so the allocation enters the entries only through frequency moments at zero lag and at the delay separation; Appendix~\ref{app:centering} sets out the conditions and the invariance of the bound to this choice.

If the occupied bands were assigned different noise variances, the covariance would become block-diagonal, $\mathbf{C}=\mathrm{diag}(\sigma_{s,1}^2\mathbf{I}_{N_{s,1}},\,\sigma_{s,2}^2\mathbf{I}_{N_{s,2}})$, with $\mathbf{I}_{N_{s,i}}$ the identity matrix of the size of band $i$, and the FIM entries would follow the general complex-Gaussian form $\left[\mathbf{I}_s(\boldsymbol{\theta})\right]_{ij}=2\,\Re\{\mathbf{d}_i^H\mathbf{C}^{-1}\mathbf{d}_j\}$~\cite{kay1993fundamentals}, in which the contribution of band $i$ is weighted by $1/\sigma_{s,i}^2$.

To isolate the information about \(\boldsymbol{\tau}\), FIM in Eq.~\eqref{eq:crlb_fim_general} can also be partitioned according to the parameter ordering \(\boldsymbol{\theta}=[\boldsymbol{\tau}^T\ \boldsymbol{\alpha}_{\mathbb{R}}^T]^T\), where \(\boldsymbol{\alpha}_{\mathbb{R}}=[\alpha_{1\mathrm{Re}}\ \alpha_{1\mathrm{Im}}\ \alpha_{2\mathrm{Re}}\ \alpha_{2\mathrm{Im}}]^T\), yielding the blocks \(\mathbf{I}_{\tau\tau}\), \(\mathbf{I}_{\tau\alpha}\), \(\mathbf{I}_{\alpha\tau}\), and \(\mathbf{I}_{\alpha\alpha}\).

Then, the effective information for \(\boldsymbol{\tau}\) is the Schur complement:
\begin{equation}
  \mathbf{I}_{\mathrm{eff}}(\boldsymbol{\tau})
  =
  \mathbf{I}_{\tau\tau}
  -
  \mathbf{I}_{\tau\alpha}\,
  \mathbf{I}_{\alpha\alpha}^{-1}\,
  \mathbf{I}_{\alpha\tau}.
  \label{eq:crlb_fim_eff}
\end{equation}
The CRLB for any locally unbiased estimator \(\widehat{\boldsymbol{\tau}}\) now is:
\begin{equation}
  \mathrm{cov}\!\big(\widehat{\boldsymbol{\tau}}\big)
  \succeq
  \mathbf{I}_{\mathrm{eff}}^{-1}(\boldsymbol{\tau}).
  \label{eq:crlb_tau}
\end{equation}
In this analysis, we consider delay separation as the estimand. The Wi-Fi CSI comes from a bistatic ISAC setup in which the transmitter and receiver run on independent clocks, so the link carries an unknown timing offset that shifts all absolute delays together but leaves their differences unaffected~\cite{wu2024sensing}.

The two estimands also differ in what their bounds capture. Since the complex gains are unknown, the delay information is not the block $\mathbf{I}_{\tau\tau}$ but $\mathbf{I}_{\mathrm{eff}}(\boldsymbol{\tau})$ of Eq.~\eqref{eq:crlb_fim_eff}, whose entries $\tilde{I}_{ij}$ are derived in Appendix~\ref{app:ieff}. Being symmetric and $2\times2$, $\mathbf{I}_{\mathrm{eff}}$ inverts in closed form, so that Eq.~\eqref{eq:crlb_tau} gives:
\begin{equation}
\mathrm{var}(\hat{\tau}_1)\ \geq\ \frac{\tilde{I}_{22}}{\tilde{I}_{11}\tilde{I}_{22}-\tilde{I}_{12}^{\,2}},\qquad
\mathrm{var}(\hat{\tau}_2)\ \geq\ \frac{\tilde{I}_{11}}{\tilde{I}_{11}\tilde{I}_{22}-\tilde{I}_{12}^{\,2}},
\label{eq:crlb_individual}
\end{equation}
in each of which the coupling of the two delays, $\tilde{I}_{12}$, enters only through its square, so that both bounds grow with its magnitude and neither responds to its sign.

For the delay separation \(\Delta\tau\) from Eq.~\eqref{eq:delta_tau_def}, we can write \(\Delta\tau=\mathbf{g}^T\boldsymbol{\tau}\) with \(\mathbf{g}=[-1\ \ 1]^T\). The corresponding bound, denoted $\mathrm{CRLB}(\Delta\tau)$ and obtained from the same inverse is:
\begin{equation}
  \mathrm{var}\!\big(\widehat{\Delta\tau}\big)
  \ge
  \mathbf{g}^{T}\,\mathbf{I}_{\mathrm{eff}}^{-1}(\boldsymbol{\tau})\,\mathbf{g}
  \;=\;\frac{\tilde{I}_{11}+\tilde{I}_{22}+2\tilde{I}_{12}}
  {\tilde{I}_{11}\tilde{I}_{22}-\tilde{I}_{12}^{\,2}}.
  \label{eq:crlb_delta_tau}
\end{equation}
Since the off-diagonal entry of $\mathbf{I}_{\mathrm{eff}}^{-1}(\boldsymbol{\tau})$ is $-\tilde{I}_{12}/(\tilde{I}_{11}\tilde{I}_{22}-\tilde{I}_{12}^{2})$, this sign decides how the errors of the two delays combine. For $\tilde{I}_{12}>0$, displacing the two estimated delays in opposite directions changes $\boldsymbol{\mu}_s$ less than displacing them in the same direction, so it is more easily masked by noise. The errors are then negatively correlated, add in the separation, and raise its bound above the sum of the individual bounds. For $\tilde{I}_{12}<0$, the displacement in the same direction is the one masked by noise, so the errors are positively correlated and partly cancel in the separation, whose bound falls below that sum.

The two bounds therefore answer different questions. Those in Eq.~\eqref{eq:crlb_individual} give the precision available on each delay taken on its own, which for $\tau_1$ constitutes the metric required in absolute ranging. However, the delay separation bound cannot be obtained directly from the individual variances, since the numerator of Eq.~\eqref{eq:crlb_delta_tau} carries also the sign of the coupling term $2\tilde{I}_{12}$ that neither contains.

\begin{figure}[t]
  \centering
  \includegraphics[width=1.0\linewidth]{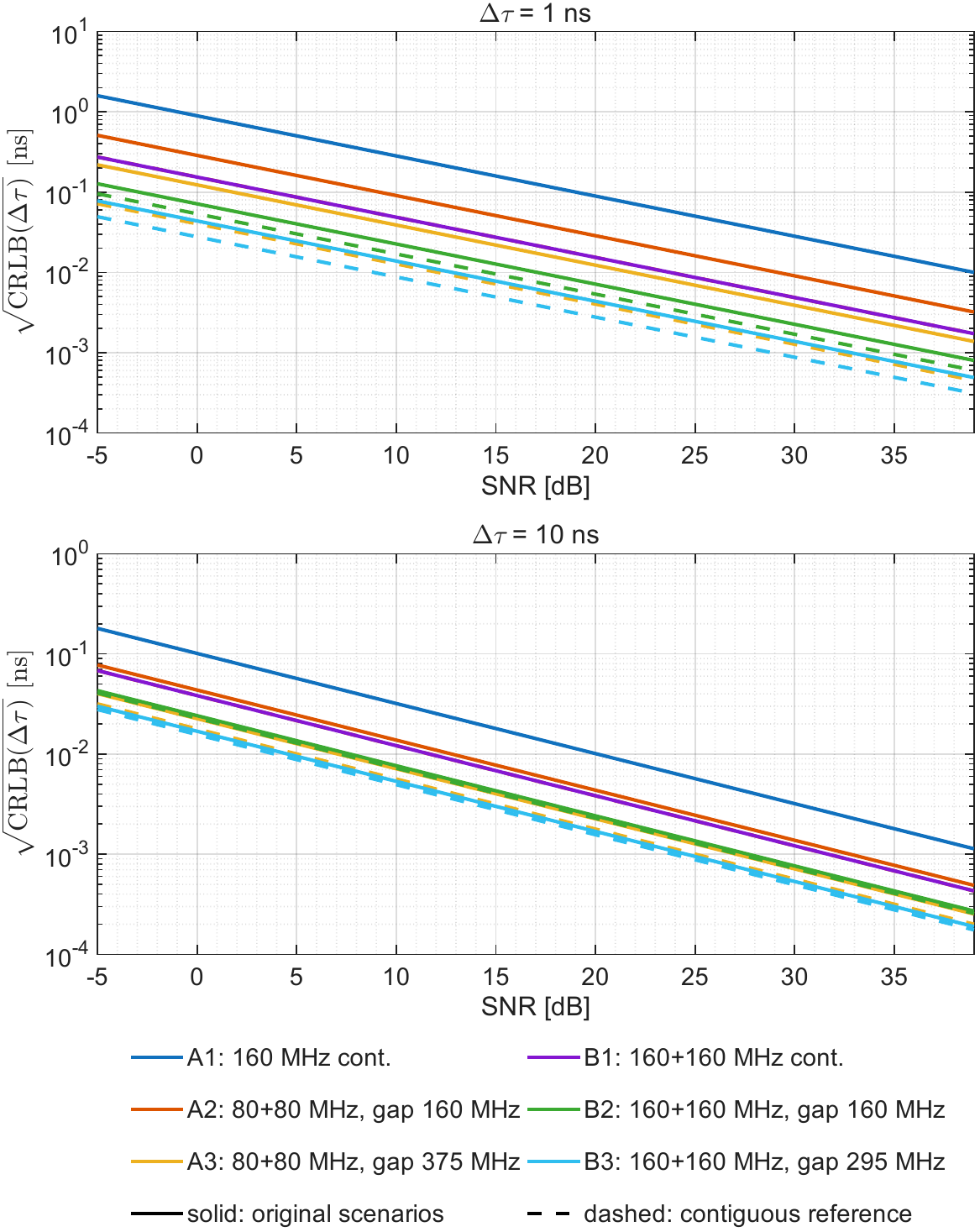}
  \caption{Square-root CRLB of $\Delta\tau$ versus SNR for scenarios A1-A3 and B1-B3, shown for $\Delta\tau=1$~ns (top) and $\Delta\tau=10$~ns (bottom).
  Contiguous references are included for the gapped cases.}
  \label{fig:crlb_snr}
\end{figure}

\subsection{CRLB Results}
\label{subsec:crlb_results}

Delay separation resolvability is quantified using the CRLB of Eq.~\eqref{eq:crlb_delta_tau}. Square-root CRLB represents a lower bound on the standard deviation of any unbiased estimator of the delay separation. It is defined for any relative delay $\Delta\tau$, taking smaller values when two paths separated by $\Delta\tau$ can be resolved with higher precision, and larger values when the precision is lower. All CRLB results are evaluated for the two-path channel of Section~\ref{subsec:twopath_formulation}.

We define signal-to-noise (SNR) as the average per occupied-subcarrier (noiseless) power-to-noise ratio:
\begin{equation}
\mathrm{SNR}
=
\frac{\|\boldsymbol{\mu}_s(\boldsymbol{\theta})\|_2^2/N_s}{\sigma_s^2}
=
\frac{1}{N_s}\sum_{n\in\mathcal{K}_s}\frac{|\mu_s[n]|^2}{\sigma_s^2},
\label{eq:snr_def}
\end{equation}
where $\boldsymbol{\mu}_s(\boldsymbol{\theta})$ is given in
Eqs.~\eqref{eq:crlb_mu} and \eqref{eq:crlb_mu_entry}. For a target $\mathrm{SNR}_{\mathrm{dB}}$, we set the noise variance to:
\begin{equation}
\sigma_s^2
=
\left(\|\boldsymbol{\mu}_s(\boldsymbol{\theta}_0)\|_2^2/N_s\right)\big/10^{\mathrm{SNR}_{\mathrm{dB}}/10},
\label{eq:sigma_from_snr}
\end{equation}
where $\boldsymbol{\theta}_0$ is the true parameter vector. As the CRLB bounds locally unbiased estimators, FIM and the resulting bound are evaluated at $\boldsymbol{\theta}_0$, which also fixes the noiseless mean power used to calibrate $\sigma_s^2$. The bound is therefore a benchmark for a known channel state rather than a prediction for measured data. In practice, $\boldsymbol{\theta}_0$ and $\sigma_s^2$ are unknown and must be estimated, so a bound evaluated from these estimates inherits their errors.
Since the noise variance $\sigma_s^2$ is common across occupied subcarriers, allocation-dependent CRLB differences arise from how the allocation changes the mean model and the observed subcarrier set via $a_s[n]$ and $f[n]$, rather than from unequal per-subcarrier measurement reliability.

\textit{SNR Scaling and Aperture Advantage:} Fig.~\ref{fig:crlb_snr} shows $\sqrt{\mathrm{CRLB}(\Delta\tau)}$ versus SNR for scenario groups A and B at two representative separations, $\Delta\tau\in\{1~\mathrm{ns},\,10~\mathrm{ns}\}$. These correspond to excess path lengths \(\Delta d=c_0\Delta\tau\), with \(c_0\) the speed of light, so \(\Delta\tau=1~\mathrm{ns}\) maps to \(\Delta d\approx 0.30~\mathrm{m}\) and \(\Delta\tau=10~\mathrm{ns}\) maps to \(\Delta d\approx 3~\mathrm{m}\), which is realistic for indoor environments. 

We first compare, within each group, cases with the same occupied bandwidth: the contiguous baseline (A1/B1) and the gapped allocations (A2-A3/B2-B3), which use the same number of subcarriers but span a larger aperture. Across both delay separations, the gapped allocations achieve lower bounds, indicating that the dominant effect is the increased effective frequency aperture.
This follows directly from the delay-delay block of FIM: the diagonal terms $I_{11}$ and $I_{22}$ scale with the quadratic frequency sum $\sum_{n\in\mathcal{K}_s}|a_s[n]|^2 f[n]^2$, and this scaling carries into the effective delay information $\mathbf{I}_{\mathrm{eff}}$ and enters the corresponding CRLB in 
Eq.~\eqref{eq:crlb_delta_tau}.

\begin{figure}[t]
  \centering
  \begin{subfigure}[b]{\linewidth}
    \centering
    \includegraphics[width=0.9\linewidth]{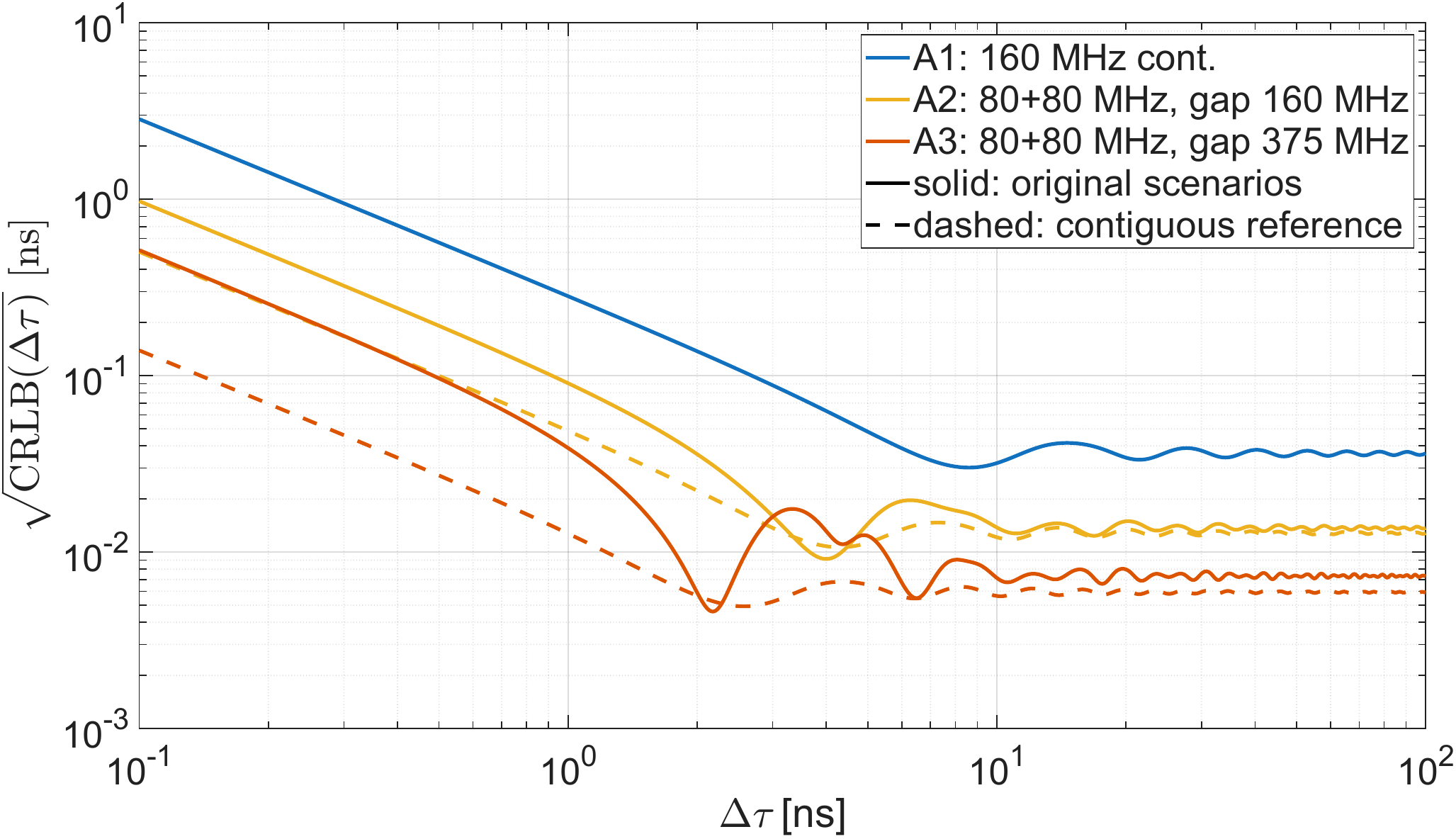}
     \caption{Group A scenarios.}
    \label{fig:crlb_dt_4a}
  \end{subfigure}
  \vspace{0.8em} 
  \begin{subfigure}[b]{\linewidth}
    \centering
    \includegraphics[width=0.9\linewidth]{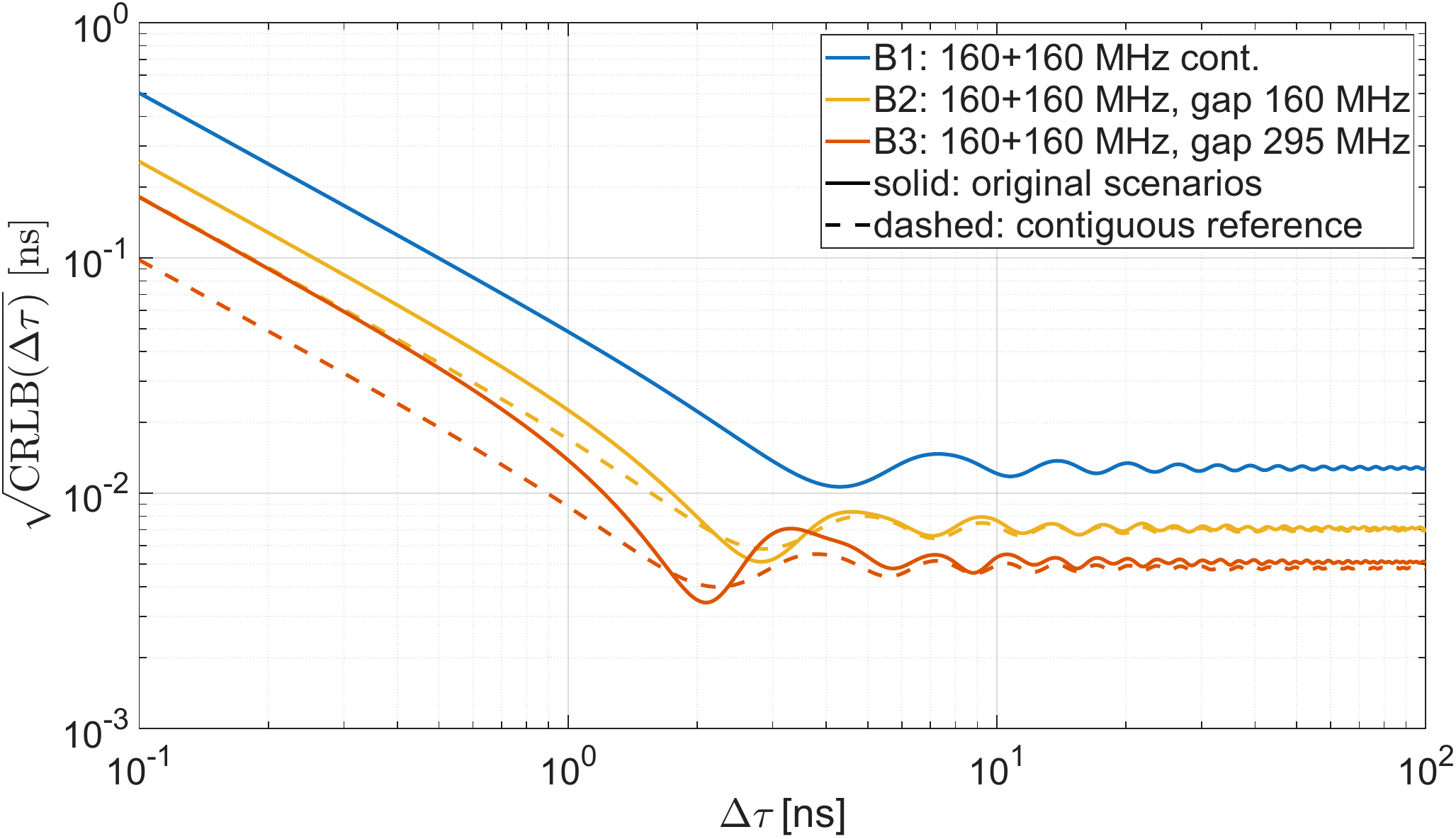}
         \caption{Group B scenarios.}
    \label{fig:crlb_dt_4b}
  \end{subfigure}
  \caption{Square-root CRLB of $\Delta\tau$ versus $\Delta\tau$ (ns) at SNR~=~10~dB.
  Contiguous references are included for the gapped cases.}
  \label{fig:crlb_dt_4}
\end{figure}

To quantify fragmentation effects, each gapped allocation (solid) is also compared to a hypothetical contiguous reference (dashed) defined over the same overall aperture. This reference assumes an OFDM subcarrier grid with the standard subcarrier spacing $\Delta f = 78.125$~kHz spanning the full aperture. Relative to this reference, the gapped allocations exhibit a higher bound. Since all FIM entries in Eq.~\eqref{eq:crlb_fim_general} scale with $1/\sigma_s^2$, this offset does not depend on the SNR and is determined by the spectral occupancy alone; it is, however, more pronounced at $\Delta\tau=1$~ns than at $\Delta\tau=10$~ns. This separation dependence motivates the $\Delta\tau$ sweep in Fig.~\ref{fig:crlb_dt_4}.

\textit{Separation Sweep and Gap Penalty:} Fig.~\ref{fig:crlb_dt_4} shows $\sqrt{\mathrm{CRLB}(\Delta\tau)}$ as a function of $\Delta\tau$ at SNR~$=10$~dB. 
The advantage of a larger frequency aperture is evident from the lower overall bound levels achieved by the allocations with a larger aperture. However, comparing each gapped allocation (solid) with its hypothetical contiguous reference over the same aperture (dashed) isolates the effect of the internal gap. For separations smaller than $1$~ns, the gapped curves exhibit an approximately constant gap-induced penalty relative to the corresponding contiguous (dashed) reference. At larger separations, the gapped scenarios first develop pronounced oscillations and, for sufficiently large $\Delta\tau$, i.e. separations bigger than $10$~ns, approach nearly the same square-root CRLB levels as the aperture-matched contiguous reference (dashed).

\textit{Asymptotic Limits at Small and Large Separations:} The two ends of this comparison are governed by different properties of the occupied subcarrier set. At small separations the two paths produce nearly the same phase progression across frequency, and the Schur complement in Eq.~\eqref{eq:crlb_fim_eff} removes almost all of their delay information. What remains grows with $\Delta\tau^{2}$ and with $\mathcal{V}_s$, yielding:
\begin{equation}
\begin{aligned}
\mathrm{CRLB}(\Delta\tau)\;&\xrightarrow[\;\Delta\tau\to0\;]{}\;
\frac{\kappa(\alpha_1,\alpha_2)\,\sigma_s^{2}}{(2\pi)^{4}\,\Delta\tau^{2}\,\mathcal{V}_s},\\[4pt]
\mathcal{V}_s\;&=\!\!\sum_{n\in\mathcal{K}_s}\!\! w[n]\big(f[n]^{2}-\overline{f^{2}}\big)^{2},
\end{aligned}
\label{eq:crlb_close}
\end{equation}
with $w[n]=|a_s[n]|^{2}$, $\overline{f^{2}}=\sum_{n\in\mathcal{K}_s}w[n]f[n]^{2}\big/\sum_{n\in\mathcal{K}_s}w[n]$ the mask-weighted mean of $f[n]^{2}$, and $\kappa(\alpha_1,\alpha_2)$ a gain-dependent factor common to all allocations. Both are derived in Appendix~\ref{app:close_sep}. Written this way, $\mathcal{V}_s$ is a mask-weighted measure of how much $f[n]^{2}$ varies over the occupied subcarriers. A gapped allocation groups its subcarriers into two subbands away from the band centroid and lacks those with the smallest $f[n]^{2}$, so $\mathcal{V}_s$ is smaller than for the aperture-matched reference.

At large delay separations the same grouping acts the other way. The coupling has decayed, the two delays are effectively estimated independently, and the bound is governed by the quadratic frequency sum $\sum_{n\in\mathcal{K}_s}|a_s[n]|^{2}f[n]^{2}$, which rewards subcarriers placed far from the band centroid rather than spread across it. This advantage compensates for the smaller number of occupied subcarriers, and the gapped curves approach the aperture-matched reference. The same grouping of subcarriers therefore penalizes closely spaced paths and favors well separated ones, through two different properties of one occupied set.

\textit{Intermediate Oscillations and Subband Decomposition:} Between these limits the coupling is neither absorbed nor negligible. Together with the information that each delay loses to the unknown gains, it produces the oscillations observed in Fig.~\ref{fig:crlb_dt_4}. The coupling $\tilde{I}_{12}$ carries a coherent sum over the occupied subcarriers:
\begin{equation}
\sum_{n\in\mathcal{K}_s}w[n]\,f[n]^{2}\,e^{-j2\pi f[n]\Delta\tau}
\label{eq:coupling_sum}
\end{equation}
in which each subcarrier contributes a phasor whose angle advances with frequency at the rate $2\pi\Delta\tau$. Across one subband of bandwidth $B_{\mathrm{sb}}$ these angles spread by only $2\pi B_{\mathrm{sb}}\Delta\tau$, so while $\Delta\tau\ll1/B_{\mathrm{sb}}$ the subband acts as a single phasor at the angle set by its center frequency, and a two-subband allocation reduces to two such phasors differing by $2\pi\Delta f_c\Delta\tau$. With the two subbands at $\mp\Delta f_c/2$, their contributions are complex conjugates, so Eq.~\eqref{eq:coupling_sum} is real and separates into an envelope and an oscillatory term:

\begin{multline}
\label{eq:coupling_fringe}
\sum_{n\in\mathcal{K}_s}w[n]\,f[n]^{2}e^{-j2\pi f[n]\Delta\tau} = {} \\
=2\,\big|S(\Delta\tau)\big|\,
\cos\!\big(\pi\Delta f_c\Delta\tau-\Phi(\Delta\tau)\big),
\end{multline}
where:
\begin{equation}
S(\Delta\tau)=\!\!\sum_{n\in\mathcal{K}_{s,2}}\!\! w[n]f[n]^{2}\,
e^{-j2\pi\left(f[n]-\Delta f_c/2\right)\Delta\tau},
\qquad \Phi=\arg S,
\label{eq:subband_S}
\end{equation}
is the upper subband referred to its own center frequency. The two factors carry the two scales of the allocation. Since $S$ spans only the subband bandwidth, $|S|$ varies on the scale $1/B_{\mathrm{sb}}$ and sets the envelope, while the cosine alternates on the scale $1/\Delta f_c$ and sets the oscillation: the sum changes sign wherever the argument crosses an odd multiple of
$\pi/2$, that is near:
\begin{equation}
\Delta\tau_k\;\approx\;\frac{k+\tfrac12}{\Delta f_c}
+\frac{\Phi}{\pi\Delta f_c},\qquad k=0,1,2,\dots,
\label{eq:sign_changes}
\end{equation}
and the bound inherits that period, consistent
with~\cite{wan2024fundamental}. The slow drift of $\Phi$ displaces the crossings from the nominal grid $(k+\tfrac12)/\Delta f_c$ and lets them migrate with $\Delta\tau$. Because $\Phi$ grows with $\Delta\tau$, the oscillation period departs from $1/\Delta f_c$, decreasing as $\Delta\tau$ increases.
In our scenarios, $\Delta f_c\in\{240,\,455,\,320,\,455\}$~MHz for A2, A3, B2, and B3, respectively, with spacings $1/\Delta f_c \approx \{4.2,\,2.2,\,3.1,\,2.2\}$~ns around which pronounced oscillations occur.

Finally, we emphasize that the CRLB is a local bound. It is determined by FIM at the true parameter and therefore captures local sensitivity and inter-parameter coupling (including the coupling of the two MPCs in the two-path model). However, it characterizes precision only within the local region of the true delay parameter, and does not account for global errors such as ambiguous peak selection.

\begin{figure}[t]
  \centering
  \includegraphics[width=0.9\linewidth]{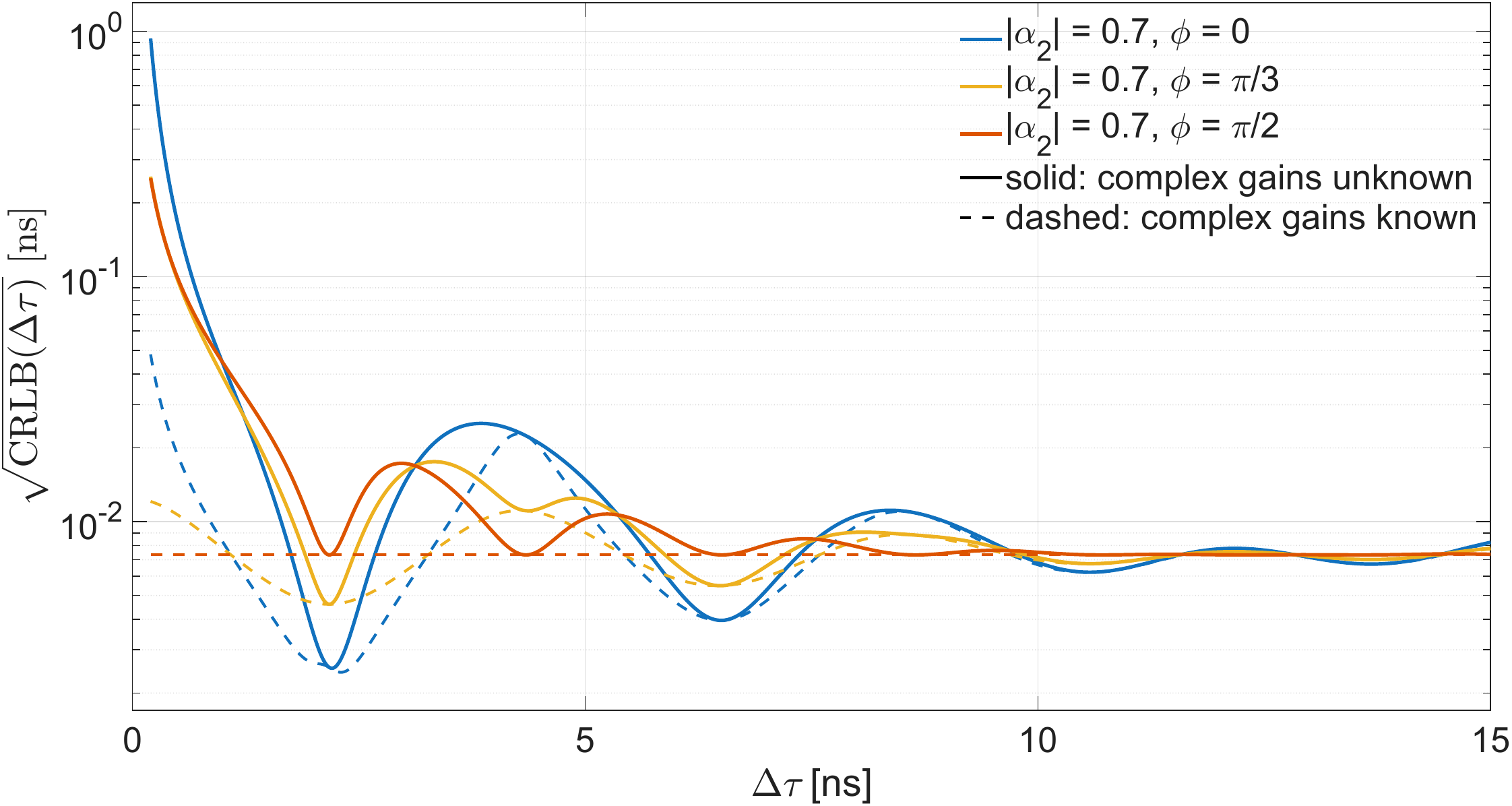}
    \caption{Bound for A3 at $\mathrm{SNR}=10$~dB for three relative phases of the second MPC, $|\alpha_2|=0.7$, $\phi\in\{0,\pi/3,\pi/2\}$. Solid: gains unknown; dashed: gains known.}
  \label{fig:ripple_phase}
\end{figure}

\subsection{Phase-State Dependence and Standard Mask Evaluation}
\label{subsec:ripple}

\textit{Impact of Relative Gain Phase and Unknown Gains:} For the symmetric, frequency centred allocations in Table~\ref{tab:scenarios} the sum in Eq.~\eqref{eq:coupling_sum} is real, so the coupling enters the delay block scaled by $\Re\{\alpha_2\}=|\alpha_2|\cos\phi$, which vanishes at $\phi=\pi/2$. With the complex gains known the delays then decouple at every delay separation and the bound is flat (Fig.~\ref{fig:ripple_phase}, dashed). The oscillations nevertheless persist for the same channel state when the gains are unknown (solid), since the Schur complement of the nuisance gains, while leaving the coupling zero, removes part of the information on each delay itself, by an amount that depends on the separation but not on $\phi$. The delay separations at which the bound oscillates are therefore set by the allocation, and their depth by the MPCs present.

\textit{Evaluation of the Rectangular Mask Idealization:} The rectangular mask in Section~\ref{subsec:observations} differs from the standard-compliant mask in that it also includes the guard and null subcarriers of the IEEE~802.11ax subcarrier plan. Including these subcarriers reduces $\sqrt{\mathrm{CRLB}(\Delta\tau)}$ at all separations, by at most $3.5\%$ in Group~A and $2.7\%$ in Group~B. The rectangular mask therefore slightly overestimates the achievable precision. Most of this reduction is due to the guard subcarriers at the channel edges, since the bound is most sensitive to the extreme values of $f[n]^2$. The reduction is largest in Group~A at small separations. In A2 and A3, the guard subcarriers occupy a larger portion of the 80~MHz channels than of the 160~MHz channels in Group~B, and in A1 they are located at the edges of the whole aperture. Since all scenarios are affected to a similar extent, the comparisons between them and the conclusions of this section also apply to the standard-compliant mask.

\section{Delay-Domain Sidelobe Structure and Matched-Filter Offset}
\label{sec:Sidelobes}

Non-contiguous allocations reshape the delay-domain behavior beyond what is captured by the CRLB alone. The CRLB provides a local variance bound under the assumed model, describing precision when the estimator localizes the true peak. However, under gapped spectrum, the occupied subcarriers and their standard-defined weights act as an effective spectral window whose delay-domain response exhibits deterministic sidelobes. These gap-induced sidelobes can produce competing peaks and delay separation-selective ambiguities, so practical estimators that rely on correlation or peak picking may exhibit outliers and threshold effects even when the CRLB remains small.
This section therefore characterizes that response: (i) for a single MPC in Section~\ref{subsec:singlePath}, where the sidelobe structure is fixed by the subband spacing; (ii) for two MPCs in Section~\ref{subsec:twoPath}, where it displaces the peaks of a delay scan by an amount obtained in closed form; and (iii) against the bound in Section~\ref{subsec:leakage}, where the bound, this offset and the sidelobe level are related through the same response; and (iv) for more than two MPCs in Section~\ref{subsec:Lpath}.

\subsection{Single-Path Delay Response and Sidelobe Characteristics}
\label{subsec:singlePath}

\textit{Delay Autocorrelation Function (ACF):} In commodity Wi-Fi CSI processing, correlation-based delay estimation is the default approach, since most sensing and ranging pipelines start with an inverse DFT of the CSI followed by a first-peak or strongest-peak rule, which requires neither a model order nor an initialization. Methods that estimate paths jointly, or through subspace decomposition, do not perform such a scan, but are conditioned by the same underlying quantity, which appears as the inner product between the responses to two candidate delays. That quantity is the ACF of the multiband spectral window, so the effect of aggregating multiple bands on delay estimation is captured entirely by this one function, regardless of which of these estimators is used.

Testing a candidate delay $\tau$ correlates the observed CFR with the
mask-weighted single-path template $a_s[n]\,e^{-j2\pi f[n]\tau}$:
\begin{equation}
\label{eq:corr}
\begin{split}
c_s(\tau) &= \sum_{n\in\mathcal{K}_s}\big(a_s[n]\,e^{-j2\pi f[n]\tau}\big)^{*} a_s[n]\,H\big(f[n]\big) = \\
&= \sum_{n\in\mathcal{K}_s} |a_s[n]|^{2}\,H\big(f[n]\big)\,e^{j2\pi f[n]\tau},
\end{split}
\end{equation}
where the mask power weights $|a_s[n]|^{2}$ are produced by the
conjugate template. For a single path, $H(f)=\alpha e^{-j2\pi f\tau_0}$:
\begin{equation}
\label{eq:corr_single}
c_s(\tau)=\alpha\!\!\sum_{n\in\mathcal{K}_s}\!\! |a_s[n]|^{2}\,e^{j2\pi f[n](\tau-\tau_0)} ,
\end{equation}
so the complex gain, which also carries the unknown receiver phase,
scales the correlation but does not shape it, and cancels in
$|c_s(\tau)|/|c_s(\tau_0)|$. What remains, written at any time offset
$u=\tau-\tau_0$ of the tested delay from the path delay, is:
\begin{equation}
\label{eq:gs}
g_s(u)=\!\!\sum_{n\in\mathcal{K}_s}\!\! |a_s[n]|^{2}\,e^{j2\pi f[n]u},
\end{equation}
so that $c_s(\tau)=\alpha\,g_s(\tau-\tau_0)$. Being the inverse discrete Fourier transform
of the mask power over the occupied subcarriers, $g_s$ is the ACF of the multiband spectral window \cite{stoica2005spectral}.

\textit{Two-Subband Delay Decomposition:} To explain the oscillatory sidelobe structure induced by spectral gaps, it is convenient to rewrite $g_s(u)$ as the coherent sum of the two occupied subbands. The mask splits as in Eq.~\eqref{eq:mask_decomposition}, with $a_{s,i}[n]$ supported on $\mathcal{K}_{s,i}$ around $f_{c,i}$, and since the ACF in Eq.~\eqref{eq:gs} weights each occupied subcarrier by the mask power, subband $i$ contributes with the weights $|a_{s,i}[n]|^2$.

To factor out the subband-center phase term and isolate the within-subband shape, we define the basebanded response of each subband $i\in\{1,2\}$ as:

\begin{equation}
G_i(u)= \sum_{n\in \mathcal{K}_{s,i}} 
|a_{s,i}[n]|^2\,
e^{j2\pi\bigl(f[n]-f_{c,i}\bigr)u}.
\label{eq:Gi_def}
\end{equation}
Then the overall response can be written as:
\begin{equation}
g_s(u)=e^{j2\pi f_{c,1}u}G_1(u)+e^{j2\pi f_{c,2}u}G_2(u).
\label{eq:gs_twosubband}
\end{equation}

Since Eq.~\eqref{eq:gs_twosubband} is complex-valued and the deterministic
sidelobe pattern is governed by its magnitude, we analyze $|g_s(u)|$. Each
subband response carries a phase of its own, distinct from the subband-centre
phase already factored out in Eq.~\eqref{eq:Gi_def}; writing $G_i(u)=|G_i(u)|e^{j\phi_i(u)}$,
the magnitude of Eq.~\eqref{eq:gs_twosubband} can be written as:
\begin{equation}
\begin{aligned}
|g_s(u)| = \Bigl[\,&|G_1(u)|^2+|G_2(u)|^2\\
&{}+2\,|G_1(u)|\,|G_2(u)|\,
\cos\bigl(2\pi\Delta f_c\,u+\Delta\phi(u)\bigr)\Bigr]^{1/2}\!,
\end{aligned}
\label{eq:gs_mag}
\end{equation}
where $\Delta\phi(u)=\phi_2(u)-\phi_1(u)$ is the difference between the phases of the two subband responses, and $\Delta f_c$, defined in Eq.~\eqref{eq:dfc}, is the spacing between the subband centre frequencies. The two enter Eq.~\eqref{eq:gs_mag} together, but on different scales: each $G_i$ spans only the subband bandwidth $B_{\mathrm{sb}}$, so $\Delta\phi$ varies on the scale $1/B_{\mathrm{sb}}$, whereas the centre-spacing term advances at $2\pi\Delta f_c$ and, since $\Delta f_c > B_{\mathrm{sb}}$ for any gapped allocation, carries faster variation. The oscillation period is therefore set by $\Delta f_c$, with $\Delta\phi$ displacing its extrema rather than setting their spacing. This corresponds to the same mechanism that governs the coupling extrema in Eq.~\eqref{eq:coupling_fringe}. At the reinforcement and cancellation extremes ($\cos=\pm1$), the bracket collapses to the perfect square $(|G_1|\pm|G_2|)^2$, so $|g_s(\tau)|$ is confined to $\big||G_1|-|G_2|\big|\le|g_s(\tau)|\le|G_1|+|G_2|$, with $|G_1|+|G_2|$ and $\big||G_1|-|G_2|\big|$ as its upper and lower envelopes.

\textit{Equal-Bandwidth Sidelobe Oscillations:} In the equal-bandwidth case, where the two subbands share the same bandwidth and weighting, $|G_1|\approx|G_2|\equiv|G|$ and $\Delta\phi\approx0$, the lower envelope $\big||G_1|-|G_2|\big|\to0$ and, using $1+\cos\theta=2\cos^2(\theta/2)$, Eq.~\eqref{eq:gs_mag} reduces to:
\begin{equation}
|g_s(u)|\approx 2\,|G(u)|\,|\cos(\pi\Delta f_c\,u)|.
\label{eq:gs_sym}
\end{equation}
Thus, $|g_s(u)|$ swings between the upper envelope $2|G(u)|$ and zero, with deep minima near $u \approx (m+\tfrac{1}{2})/\Delta f_c$, $m=0,1,2,\ldots$, where the two subband contributions are approximately out of phase.  Additional near-zeros can occur due to the intrinsic null structure of $G_i(u)$ itself; thus, the observed minima reflect the combined effect of the per-subband envelope (set by $B_{\mathrm{sb}}$ and weighting) and inter-subband interference (set by $\Delta f_c$).

\begin{figure}[t]
  \centering

  \begin{subfigure}[b]{\linewidth}
    \centering
    \includegraphics[width=0.85\linewidth]{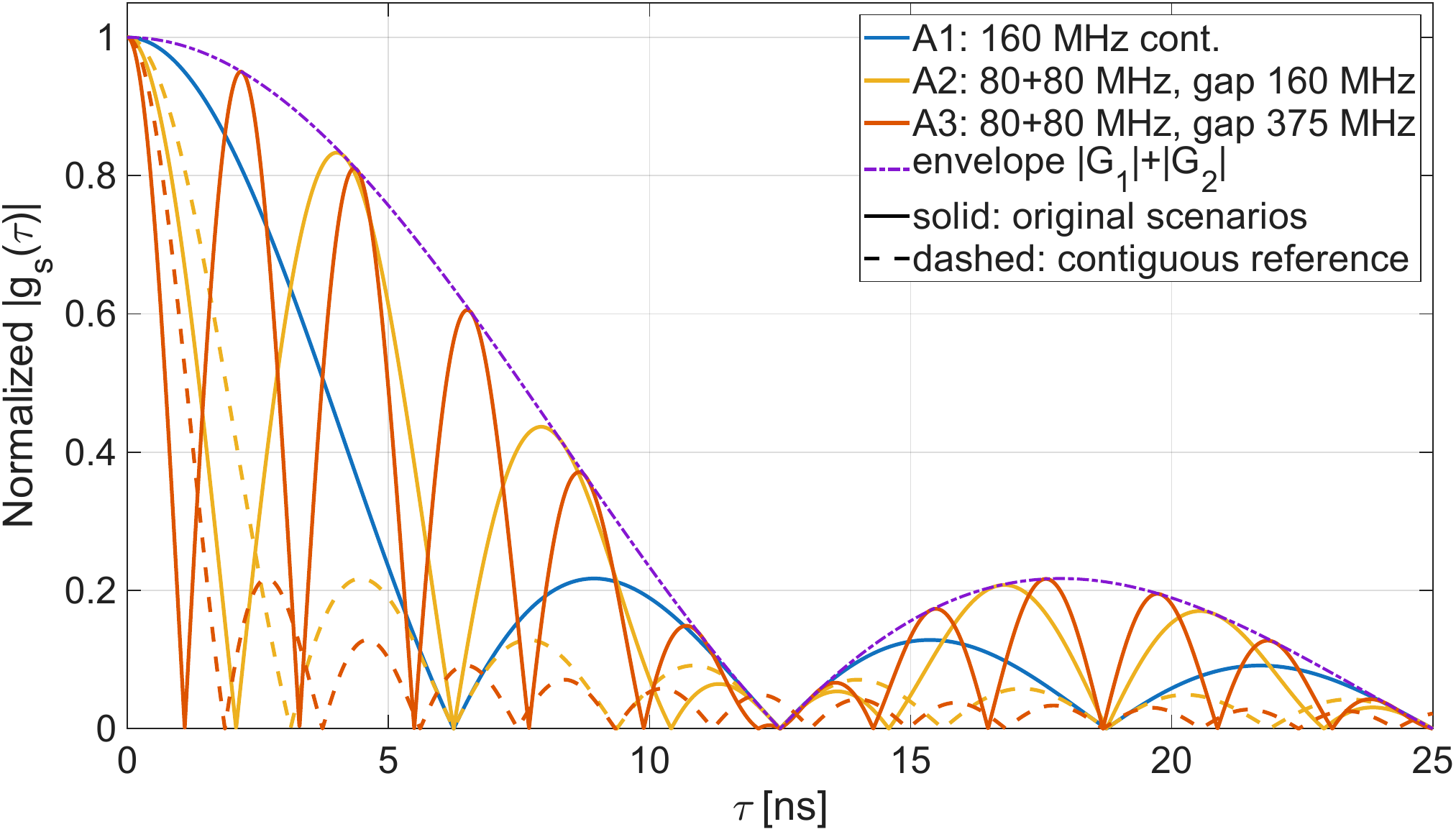}
     \caption{Group A scenarios.}
    \label{fig:gs_singlepath_a}
  \end{subfigure}

  \vspace{0.8em} 

  \begin{subfigure}[b]{\linewidth}
    \centering
    \includegraphics[width=0.85\linewidth]{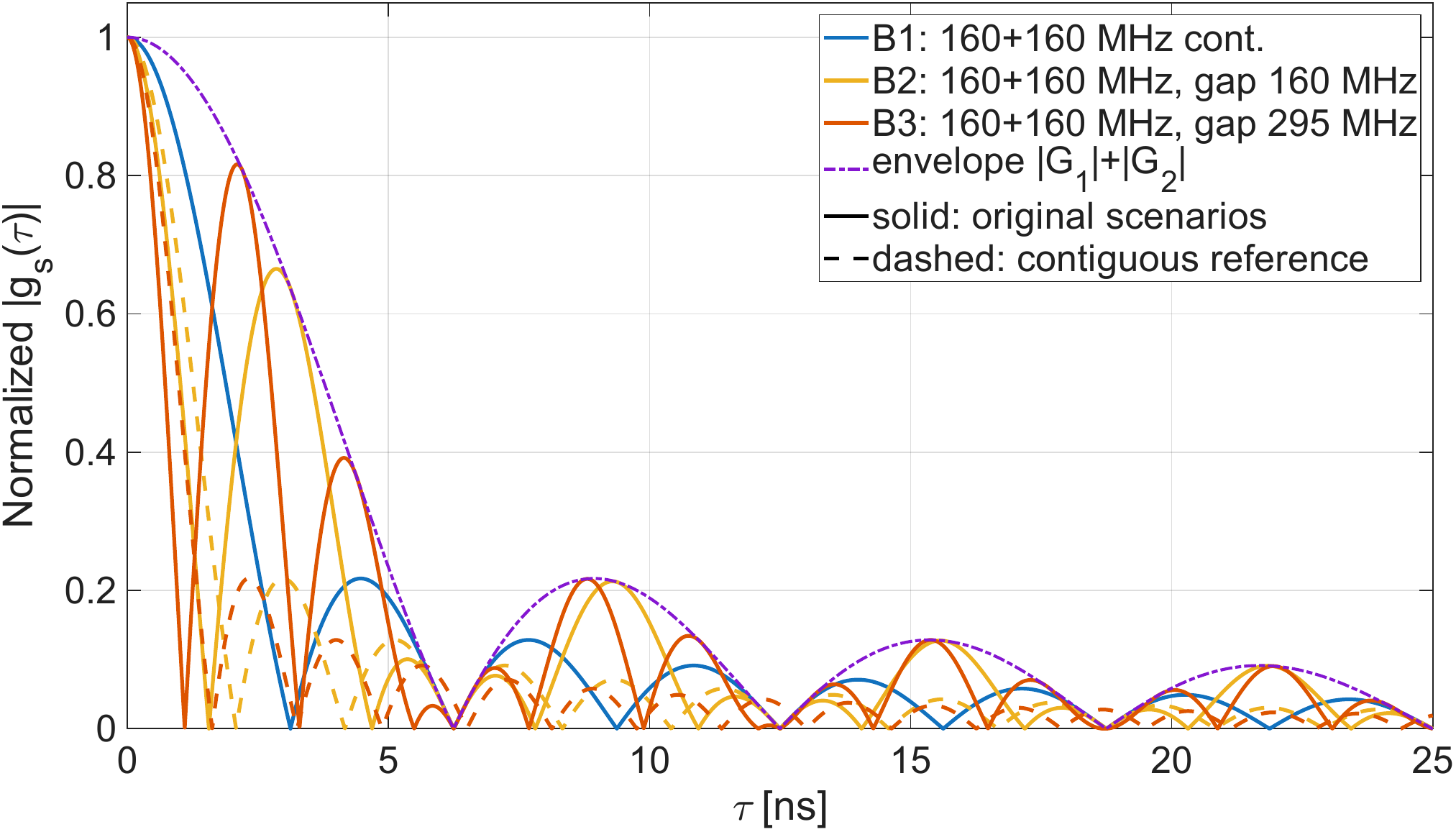}
         \caption{Group B scenarios.}
    \label{fig:gs_singlepath_b}
  \end{subfigure}

\caption{Normalized single-path delay response \(|g_s(u)|\) for scenario groups A and B, with the shared subband envelope.}

  \label{fig:sidelobes_singlepath}
\end{figure}


\textit{Delay Profile Comparison:} Fig.~\ref{fig:sidelobes_singlepath} shows the normalized magnitude of the single-path delay response $|g_s(u)|$ for the allocations in Groups~A and~B. Solid curves correspond to the original allocations, while dashed curves show the contiguous reference variants for the gapped cases. In both groups, contiguous allocations produce a single dominant mainlobe followed by smoothly decaying sidelobes, whereas the gapped allocations of each group oscillate beneath their shared subband envelope $|G_1|+|G_2|$ (dash-dotted). Because each $G_i(u)$ depends only on the subband bandwidth and within-band weighting, not on the gap width, scenarios built from the same channel bandwidth (A2 and A3 at $80$~MHz, B2 and B3 at $160$~MHz) share this envelope and differ only in the oscillation rate set by $\Delta f_c$. Comparing each gapped allocation to its contiguous reference shows that the internal gap narrows the mainlobe, but raises the sidelobes and shifts the locations of the local maxima and minima. For illustration, consider scenario A2, for which $\Delta f_c=240$~MHz and $B_{\mathrm{sb}}=80$~MHz. This yields gap-induced minima near $\tau_{0,m}\approx\{2.08,\,6.25,\,10.42,\,14.58,\,18.75,\,\ldots\}$~ns. In addition, the main envelope produces minima at roughly integer multiples of $1/B_{\mathrm{sb}}$, which for A2 corresponds to $\{12.5,\,25,\,\ldots\}$~ns.

\begin{figure}[t]
  \centering
  \includegraphics[width=0.85\linewidth]{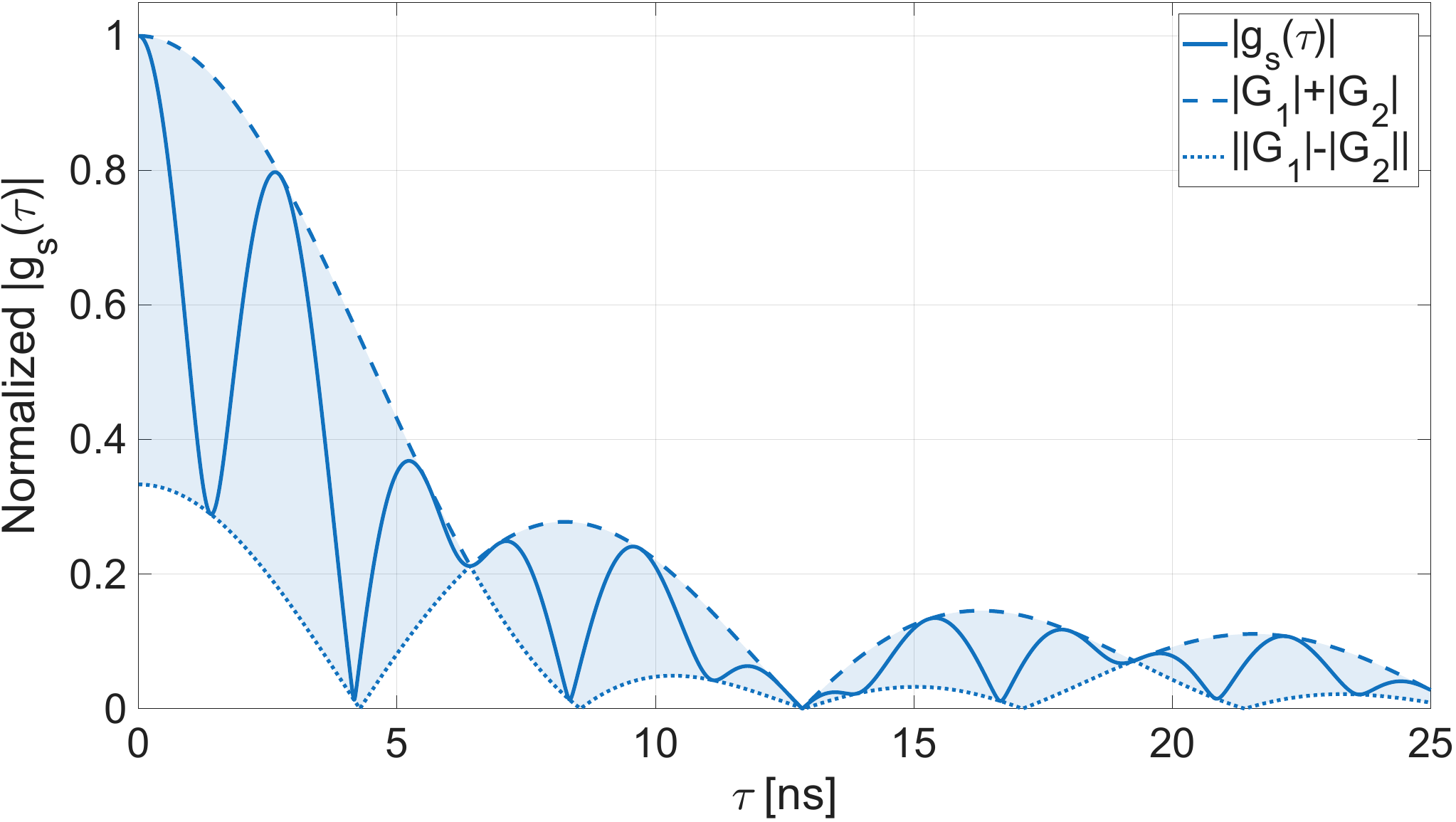}
  \caption{Single-path delay response $|g_s(u)|$ for an asymmetric allocation ($80$~MHz $+$ $160$~MHz, aperture of B2), with envelopes $|G_1|+|G_2|$ (dashed) and $\big||G_1|-|G_2|\big|$ (dotted).}
  \label{fig:gs_envelope}
\end{figure}

\begin{figure}[t]
  \centering
  \begin{subfigure}[b]{\linewidth}
    \centering
    \includegraphics[width=0.9\linewidth]{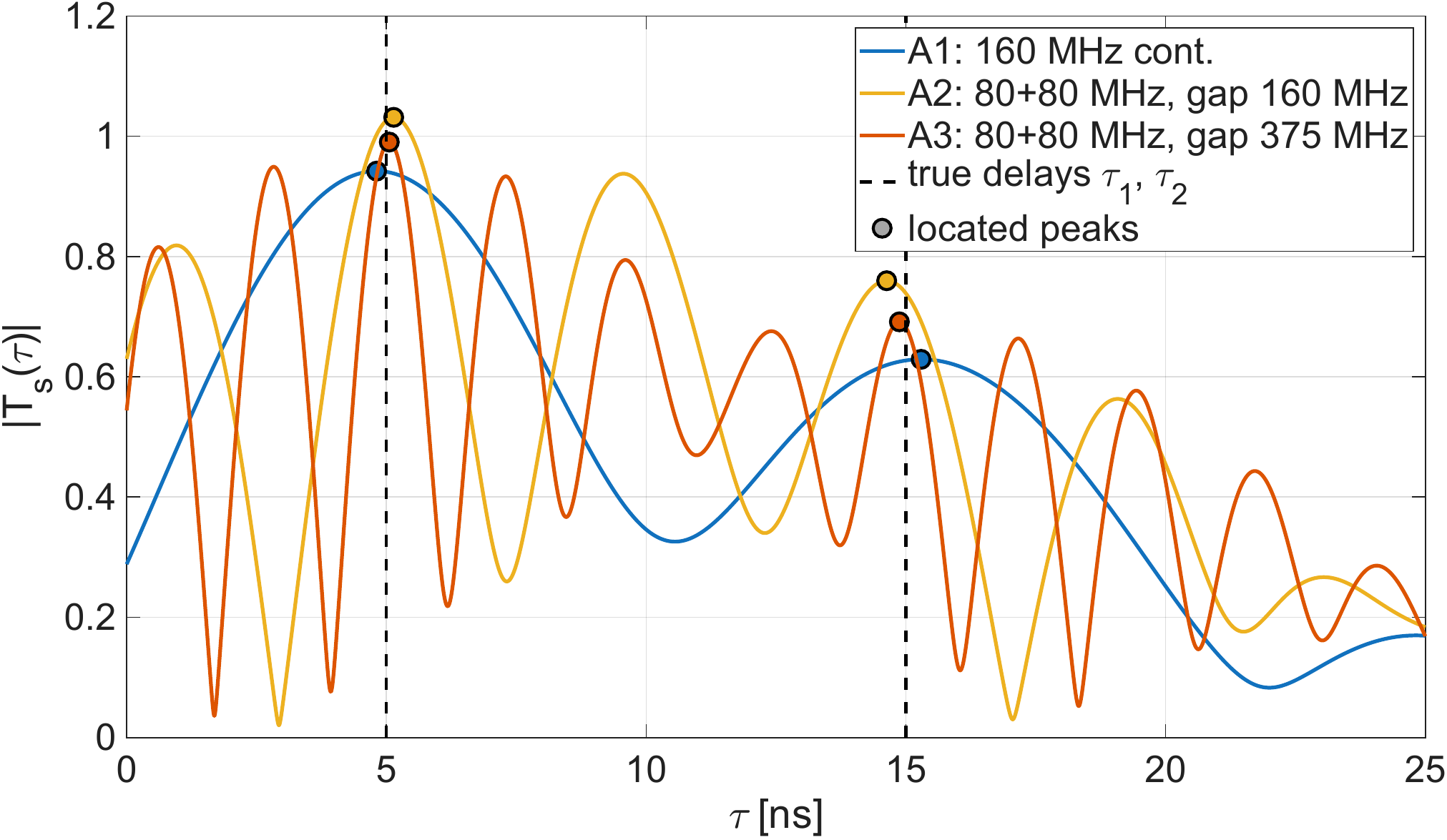}
    \caption{Group A scenarios (\(\tau_1=5\) ns, \(\tau_2=15\) ns).}
    \label{fig:sidelobes_twopath_a}
  \end{subfigure}
  \vspace{0.8em}
  \begin{subfigure}[b]{\linewidth}
    \centering
    \includegraphics[width=0.9\linewidth]{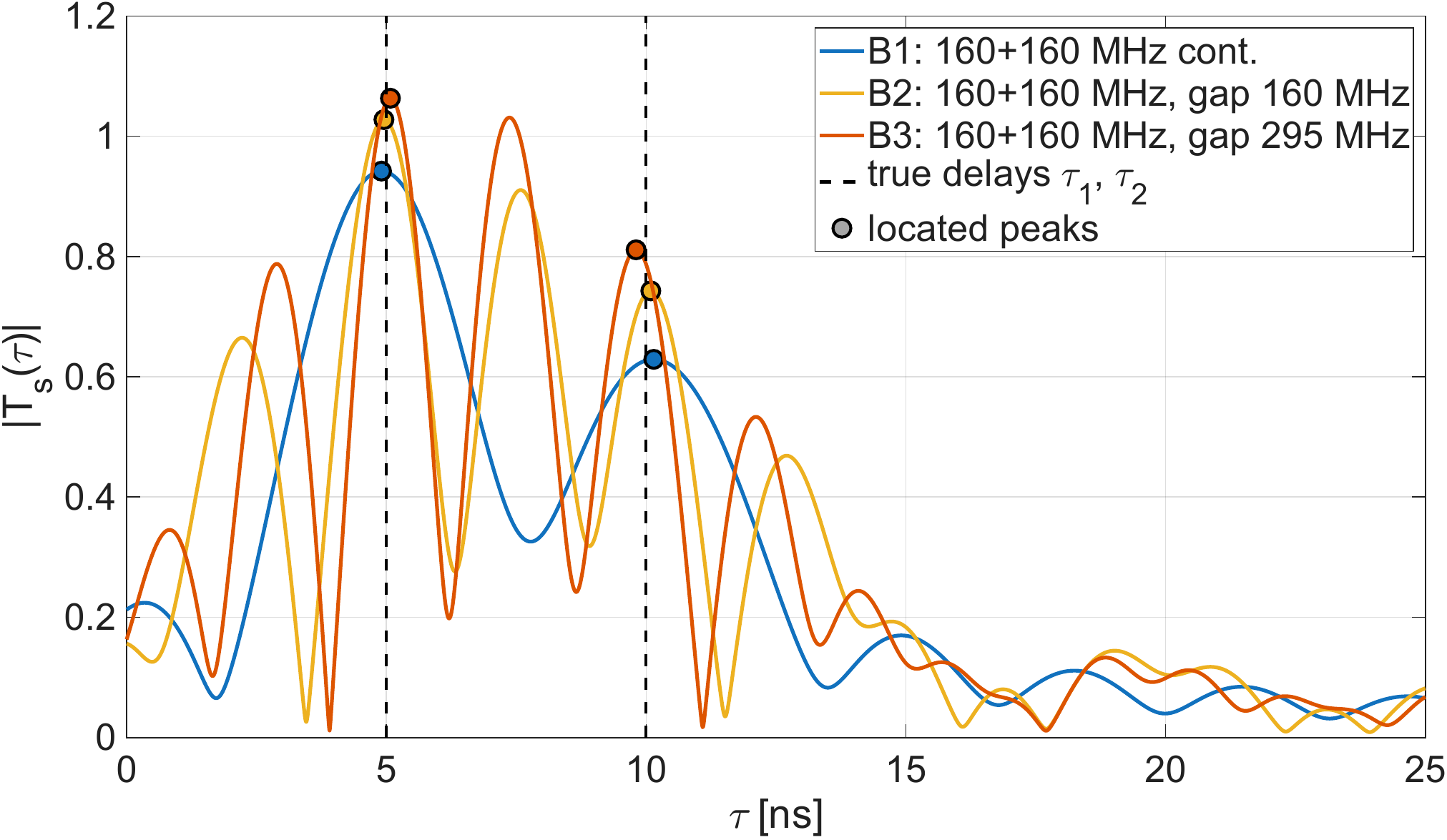}
    \caption{Group B scenarios (\(\tau_1=5\) ns, \(\tau_2=10\) ns).}
    \label{fig:sidelobes_twopath_b}
  \end{subfigure}
  \caption{Noise-free single-path correlation output $T_s(\tau)$ of the \emph{single-path} matched filter %
  applied to the two-path channel. Vertical dashed lines mark the true delays; markers give the located peaks $\hat\tau_1,\hat\tau_2$ in Table~\ref{tab:twopath_peak_offsets}.}
  \label{fig:sidelobes_twopath}
\end{figure}

\textit{Asymmetric Subband Configurations:} When the two subbands differ in bandwidth, $|G_1|\neq|G_2|$, the lower envelope $\big||G_1|-|G_2|\big|$ no longer reaches zero, so the minima fill in rather than vanish. Fig.~\ref{fig:gs_envelope} shows this for a hypothetical $80$~MHz $+$ $160$~MHz allocation spanning the aperture of B2 scenario.

\textit{Evaluation of the Rectangular Mask Idealization:} Compared with the rectangular mask used so far, the standard-compliant mask, in which the guard and null subcarriers are unused, changes $|g_s(u)|$ by at most $2.8\%$ of its peak.

\subsection{Two-Path Delay Response and Deterministic Matched-Filter Offset}
\label{subsec:twoPath}

In practical multipath propagation environments, the received CFR constitutes a superposition of multiple delay-dependent phase progressions, causing the gap-induced sidelobe structure to directly impact practical estimators that evaluate delay candidates through peak selection.

\textit{Two-Path Correlation Scan and Observed Peak Offsets:} To illustrate this, a standard matched-filter delay scan is obtained by testing candidate delays \(\tau\). For each \(\tau\), the CFR samples on the occupied subcarriers are correlated with the conjugate of the phase progression that a single propagation path at delay \(\tau\) would induce across frequency, i.e., \(e^{+j2\pi f[n]\tau}\). Applying the correlation from Eq.~\eqref{eq:corr} to the two-path CFR in Eq.~\eqref{eq:two_path} and using Eq.~\eqref{eq:gs} term by term, the scan for scenario $s$ is:
\begin{equation}
\begin{aligned}
T_s(\tau)
&=\Big|\sum_{n\in\mathcal{K}_s}|a_s[n]|^{2}H\!\bigl(f[n]\bigr)
e^{\,j2\pi f[n]\tau}\Big|\\
&=\Big|\sum_{n\in\mathcal{K}_s}|a_s[n]|^{2}\sum_{\ell=1}^{2}\alpha_\ell
e^{-j2\pi f[n]\tau_\ell}\,e^{\,j2\pi f[n]\tau}\Big|\\
&=\Big|\sum_{\ell=1}^{2}\alpha_\ell\sum_{n\in\mathcal{K}_s}|a_s[n]|^{2}
e^{\,j2\pi f[n](\tau-\tau_\ell)}\Big|\\
&=\bigl|\alpha_1\,g_s(\tau-\tau_1)+\alpha_2\,g_s(\tau-\tau_2)\bigr|.
\end{aligned}
\label{eq:twopath_scan}
\end{equation}
The scan is the magnitude of a superposition of two shifted copies of the single-path response. Its scale is scenario-dependent: it is set by \(\alpha_1\) and \(\alpha_2\) as well as by the coherent superposition of the two contributions \(\alpha_1 g_s(\tau-\tau_1)\) and \(\alpha_2 g_s(\tau-\tau_2)\).

Fig.~\ref{fig:sidelobes_twopath} shows the normalized two-path scan \(T_s(\tau)\) for the presented scenarios, highlighting how sidelobe structure and peak contrast change with spectral gaps. The scan is evaluated without noise, with $\alpha_1=1$ and $\alpha_2=0.7e^{j\pi/3}$, and on the rectangular mask from Section~\ref{sec:Framework} (the standard-compliant mask is evaluated at the end of this subsection).
Group~A uses \((\tau_1,\tau_2)=(5,15)\) ns, and Group~B uses \((\tau_1,\tau_2)=(5,10)\) ns, i.e. separations $\Delta\tau=10$~ns and $\Delta\tau=5$~ns respectively. These values are chosen so that path separations exceed the \(1/B\) resolution limit of contiguous baseline scenarios A1 and B1 (approximately 6.25 ns and 3.13 ns respectively). This highlights how introducing a spectral gap, while using the same total bandwidth and increased aperture, changes the effective delay response, most notably the sidelobe structure and the contrast between the two peaks, relative to single-band baseline.

The results show that a larger aperture can sharpen the mainlobe (e.g., A3 versus A1), but gap-shaped sidelobes and the relative path configuration can still pull the peak away from the true delay, even when the mainlobe is sharper (e.g., A2 versus A1 for second path). Consequently, increasing aperture does not translate monotonically into improved multipath localization. Performance is jointly governed by mainlobe sharpness and the sidelobe/ambiguity structure induced by the allocation and the specific multipath geometry.

Table~\ref{tab:twopath_peak_offsets} quantifies the peak offsets visible in Fig.~\ref{fig:sidelobes_twopath}. For each path it lists the located peak $\hat\tau_\ell$, its offset from the true delay $b_\ell=\hat\tau_\ell-\tau_\ell$, and the closed-form prediction $\hat b_\ell$ of that offset, derived below. The last three columns give the delay separation the scan returns, $\widehat{\Delta\tau}_{\mathrm{MF}}=\hat\tau_2-\hat\tau_1$, its bias $b_\Delta=b_2-b_1$, and the corresponding prediction $\hat b_\Delta=\hat b_2-\hat b_1$. Peaks are located within a window $|\tau-\tau_\ell|\le W$, with $W=1$~ns. $\hat\tau_\ell$ is the position of the largest value of $|T_s(\tau)|$ on a uniform delay grid, refined by three-point parabolic interpolation. Since $W$ is smaller than both the sidelobe spacing $1/\Delta f_c$ and the delay separation, only the peak of path $\ell$ falls inside, and the offsets do not change for $W$ between $0.5$ and $1.5$~ns.

\begin{table}[t]
  \centering
    \caption{Peak offsets of the single-path matched filter (correlator) applied to the two-path channel, on a noise-free scan with a rectangular mask. Group~A: $(\tau_1,\tau_2)=(5,15)$~ns, $\Delta\tau=10$~ns; Group~B: $(5,10)$~ns, $\Delta\tau=5$~ns.}
  \label{tab:twopath_peak_offsets}
  \setlength{\tabcolsep}{2.5pt}
  \renewcommand{\arraystretch}{1.1}
  \scriptsize
  \begin{tabular}{@{}cccccccccc@{}}
    \toprule
    & \multicolumn{3}{c}{Path 1} & \multicolumn{3}{c}{Path 2}
    & \multicolumn{3}{c}{Separation}\\
    \cmidrule(lr){2-4}\cmidrule(lr){5-7}\cmidrule(lr){8-10}
    ID & $\hat\tau_1$ & $b_1$ & $\hat b_1$
       & $\hat\tau_2$ & $b_2$ & $\hat b_2$
       & $\widehat{\Delta\tau}_{\mathrm{MF}}$ & $b_\Delta$ & $\hat b_\Delta$ \\
    & [ns] & [ns] & [ns] & [ns] & [ns] & [ns]
      & [ns] & [ns] & [ns] \\
    \midrule
    A1 & 4.812 & $-0.188$ & $-0.189$ & 15.294 & $+0.294$ & $+0.295$ & 10.482 & $+0.482$ & $+0.485$ \\
    A2 & 5.140 & $+0.140$ & $+0.142$ & 14.631 & $-0.369$ & $-0.389$ & 9.490  & $-0.510$ & $-0.531$ \\
    A3 & 5.055 & $+0.055$ & $+0.056$ & 14.875 & $-0.125$ & $-0.128$ & 9.819  & $-0.181$ & $-0.183$ \\
    \midrule
    B1 & 4.906 & $-0.094$ & $-0.095$ & 10.147 & $+0.147$ & $+0.148$ & 5.241  & $+0.241$ & $+0.242$ \\
    B2 & 4.953 & $-0.047$ & $-0.048$ & 10.090 & $+0.090$ & $+0.091$ & 5.138  & $+0.138$ & $+0.139$ \\
    B3 & 5.078 & $+0.078$ & $+0.079$ & 9.807  & $-0.193$ & $-0.212$ & 4.729  & $-0.271$ & $-0.290$ \\
    \bottomrule
  \end{tabular}
\end{table}

\textit{Analytical Derivation of the Deterministic Peak Offset:} The offsets of Table~\ref{tab:twopath_peak_offsets} are deterministic, since the scan is evaluated in the noise-free regime. They are caused by the second MPC, whereby the peak of the MPC whose delay is being estimated is displaced by the response of the other. As an estimator, the scan is therefore biased by this offset, the multipath bias of~\cite{ortega2023gnss,navarro2026ofdm}. For the first MPC, this displacement has been derived to first order in the interfering gain in~\cite{dun2021design}, with the gains assumed known, and expressed in~\cite{navarro2026ofdm} through the slope of the ACF at the delay separation, so that it vanishes at the extrema of the response. Applying the same first-order step to both MPCs, as derived in Appendix~\ref{app:bias}, gives:

\begin{equation}
\hat{b}^{(1)}_{\ell}(\Delta\tau)\;=\;\frac{|\alpha_{\bar\ell}|}{|\alpha_\ell|}\,
\cos\psi_\ell\,\lambda_s(\Delta\tau)\,\frac{1}{2\pi B_{\mathrm{rms}}},
\label{eq:bias_factored}
\end{equation}
where $\hat b^{(1)}_\ell$ is the first-order prediction of the offset of path $\ell$, where $\bar\ell$ denotes the other path of the pair, and:
\begin{equation}
\begin{aligned}
\psi_\ell &= \arg\!\left(\alpha_{\bar\ell}/\alpha_\ell\right)
            + \arg g_s'(\mp\Delta\tau),\\[2pt]
\lambda_s(u) &= \frac{|g_s'(u)|}{2\pi\,g_s(0)\,B_{\mathrm{rms}}},\\[2pt]
B_{\mathrm{rms}}^{2} &= \frac{\sum_{n\in\mathcal{K}_s}|a_s[n]|^{2}f[n]^{2}}
                             {\sum_{n\in\mathcal{K}_s}|a_s[n]|^{2}}.
\end{aligned}
\label{eq:bias_defs}
\end{equation}
Here, $g_s'$ is the derivative of the single-path response from Eq.~\eqref{eq:gs}, with respect to its argument, read at $-\Delta\tau$ when the estimated component is the first and at $+\Delta\tau$ when it is the second. The phase $\psi_\ell$ therefore carries the sign of that slope together with the relative gain phase.
 The offset thus grows with the relative gain of the other MPC and with the part of its slope that is in phase with the estimated MPC, and decreases as $B_{\mathrm{rms}}$ sharpens the mainlobe of the estimated MPC. To first order, it is the same whether the gain of the estimated MPC is known or, as in the scan, left free.

\textit{Second-Order Offset:} The first order suffices for the known-gain case of the first MPC, whose expansion parameter $r_\ell=|\alpha_{\bar\ell}/\alpha_\ell|$ is below one~\cite{dun2021design,navarro2026ofdm}. The delay separation also requires the second MPC, for which $r_\ell$ is the inverse, $1.43$ against $0.7$ at $\tau_1$. Eq.~\eqref{eq:bias_factored} gives the sign of every offset in Table~\ref{tab:twopath_peak_offsets}, but its relative error, $|\hat b^{(1)}_\ell-b_\ell|/|b_\ell|$, reaches $17\%$ at $\tau_1$ and $44\%$ at $\tau_2$. Retaining the second-order terms in the same Newton step, as derived in Appendix~\ref{app:bias}, gives:
\begin{equation}
\hat b_\ell=\hat b^{(1)}_\ell\,
\frac{1+r_\ell\,\bar g_s/\cos\phi}
{1+r_\ell\cos\phi\,(\chi_s+\bar g_s)+r_\ell^{2}\,(\bar g_s\chi_s-\lambda_s^{2})},
\label{eq:bias_second}
\end{equation}
with $\bar g_s=g_s(\Delta\tau)/g_s(0)$, $\lambda_s=\lambda_s(\Delta\tau)$ and $\chi_s=g_s''(\Delta\tau)/g_s''(0)$ the level, slope and curvature of the ACF at the delay separation, that is, its value and its first and second derivatives, normalized at the peak. With the gain known, the numerator reduces to one and the denominator to its curvature term; the level, and $\lambda_s^{2}$, enter because the magnitude of the scan attributes the sidelobe of the other MPC to the gain of the estimated one. Through the level, the peak is displaced even at $\phi=\pi/2$, where Eq.~\eqref{eq:bias_factored} predicts none, as also observed numerically in~\cite{navarro2026ofdm}. Eq.~\eqref{eq:bias_second} gives the values $\hat b_\ell$ of Table~\ref{tab:twopath_peak_offsets}, with a relative error of at most $1.6\%$ at $\tau_1$ and $9.6\%$ at $\tau_2$; over all entries of the table, including the separation, the difference is at most $21$~ps. Its range of validity is discussed in Appendix~\ref{app:bias}.
Like the bound, the predictions are evaluated at the true parameters, here the gains of Section~\ref{subsec:twopath_formulation} and the true delay separation; on measured data, these must be estimated, and the predictions inherit also their errors.

\textit{Offsets Across Allocation Geometries:} The allocation enters the offset through the sharpness of the mainlobe, $B_{\mathrm{rms}}$, which a wider aperture always increases, and through the slope, level and curvature of the response of the other MPC at the delay separation, which the gap places through the period $1/\Delta f_c$. In A2 and A3 the slope at $\Delta\tau=10$~ns is nearly the same, so the larger $B_{\mathrm{rms}}$ of A3, $229$ against $122$~MHz, reduces $|b_\Delta|$ from $0.510$ to $0.181$~ns. In B3, $B_{\mathrm{rms}}$ exceeds that of B2 by $40\%$, but the response at $\Delta\tau=5$~ns is steeper and higher, and $|b_\Delta|$ doubles, from $0.138$ to $0.271$~ns, where Eq.~\eqref{eq:bias_factored} alone predicts an increase of $26\%$. Aperture therefore acts only through $B_{\mathrm{rms}}$, and whether its gain survives at a given delay separation is decided by the response there.

\textit{Combined Delay Separation Offset:} To first order, the two MPCs are pulled in opposite directions, the slope being read at $-\Delta\tau$ for the first and at $+\Delta\tau$ for the second, and in every row of Table~\ref{tab:twopath_peak_offsets} the two offsets indeed have opposite signs. The offset on the delay separation, $b_\Delta=b_2-b_1$, therefore exceeds either per-path offset in magnitude, and most of it comes from the weaker MPC, whose offset requires the second order.

\textit{Evaluation of the Rectangular Mask Idealization:} With the standard-compliant mask, the change in $b_\Delta$ is at most $60$~ps for A1 and A2, $18$~ps for A3, and less than $4$~ps for Group~B. The sign of every offset and the ranking of the allocations by $|b_\Delta|$ are unchanged. Eq.~\eqref{eq:bias_second} predicts $b_\Delta$ with a relative error of $7.5\%$ for the standard-compliant mask and $7.3\%$ for the rectangular mask.

\begin{figure}[t]
  \centering
  \includegraphics[width=0.9\linewidth]{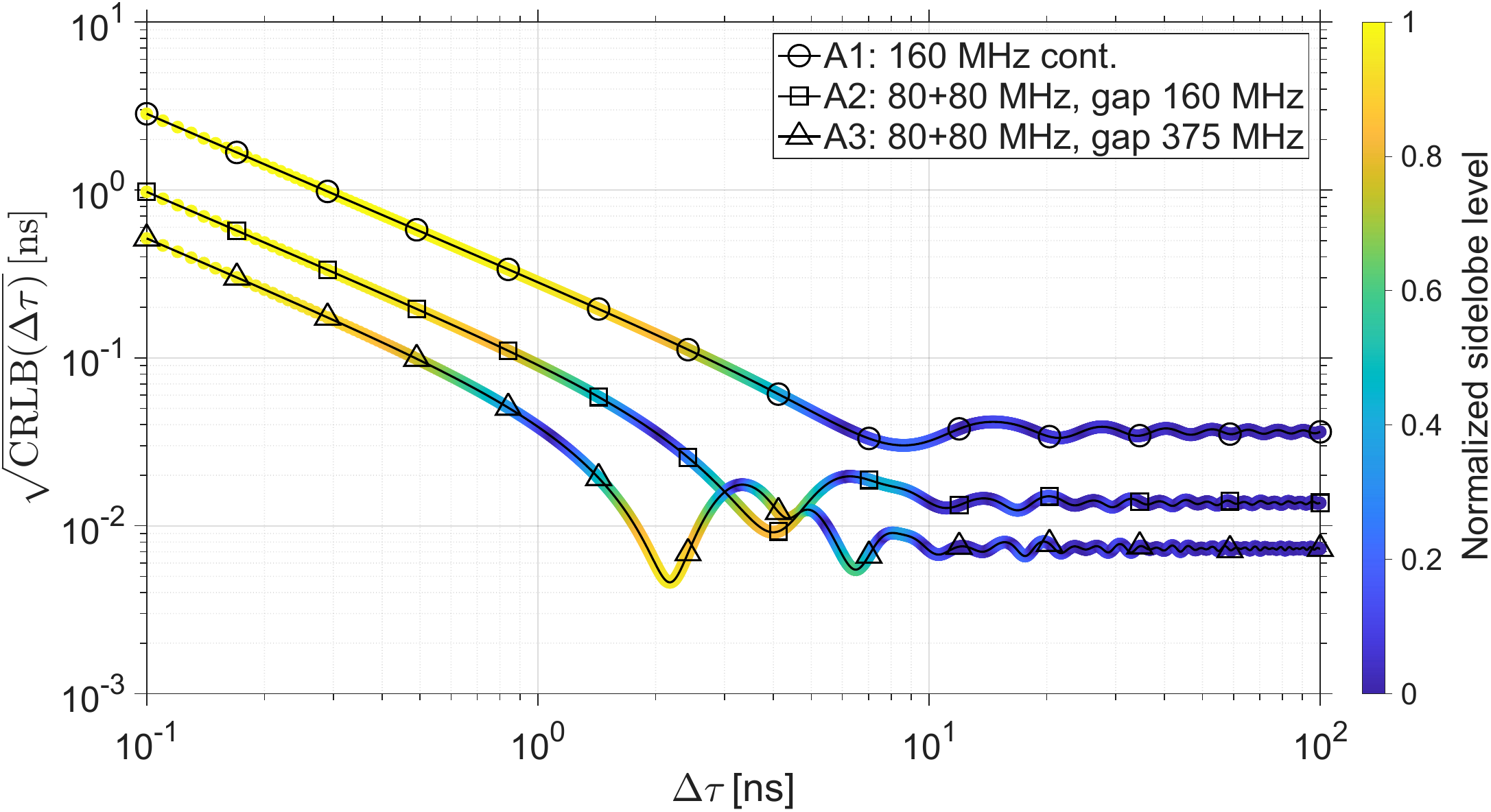}
  \caption{$\sqrt{\mathrm{CRLB}(\Delta\tau)}$ versus $\Delta\tau$ for Group~A, with normalized sidelobe level $\gamma_s(\Delta\tau)$ at SNR $=10$~dB.}
  \label{fig:leakage_crlb_AB}
\end{figure}

\subsection{\texorpdfstring{Sidelobe Structure and Practical Impact}{Sidelobe Structure and Practical Impact}}
\label{subsec:leakage}

\textit{The Bound Through the Delay Response:} 
For the first MPC with known gains, \cite{dun2021design} expresses the bound using the second derivative of the ACF at the delay separation, and uses the resulting measure of dependence as one of its constraints for band selection.
With the bands limited to the standard’s allocations, the bound is needed here to characterize what each allocation permits. With the gains unknown, it is still set by the ACF, but not by its second derivative alone.
For the symmetric masks of the allocations in Table~\ref{tab:scenarios}, this bound can be expressed through the ACF in the closed form derived in Appendix~\ref{app:ripple}:
\begin{equation}
\mathrm{CRLB}(\Delta\tau)=\frac{v_1+v_2+2\varrho_s\sqrt{v_1v_2}}{\eta_s\,(1-\varrho_s^{2})},
\label{eq:crlb_curv}
\end{equation}
where $v_\ell=\sigma_s^{2}/(2|\alpha_\ell|^{2}|g_s''(0)|)$ is the single-path bound on $\tau_\ell$~\cite{kay1993fundamentals}, and $\eta_s$ and $\varrho_s$ follow from the level $\bar g_s$, slope $\lambda_s$ and curvature $\chi_s$ of Section~\ref{subsec:twoPath} as:
\begin{equation}
\eta_s=1-\frac{\lambda_s^{2}}{1-\bar g_s^{2}},\qquad
\varrho_s=\frac{\cos\phi}{\eta_s}\Big(\chi_s-\frac{\bar g_s\,\lambda_s^{2}}{1-\bar g_s^{2}}\Big).
\label{eq:rho_curv}
\end{equation}

With the gains known, $\eta_s=1$ and $\varrho_s=\cos\phi\,\chi_s$ at every separation, and for $\phi\in\{0,\pi\}$ the bound on $\tau_1$ reduces to that of~\cite{dun2021design}. With the gains unknown, Table~\ref{tab:roles} lists, for each quantity, the parameters it couples in the FIM and what it represents in the bound and in the offset. The slope makes a shift of one delay resemble a change in the gain of the other MPC, so each delay retains only the fraction $\eta_s<1$ of its information; the same slope pulls the peaks of the scan and enters the offset in Eq.~\eqref{eq:bias_second} as a factor. The curvature correlates the two delay errors through $\varrho_s$, whose sign decides whether they add or partly cancel in the separation. The level raises the variance of the gains by $1/(1-\bar g_s^{2})$, amplifying the information loss in the bound, and raises or lowers the peak in the offset; in both, it acts only together with the slope. At the sidelobe maxima of the ACF, where the slope vanishes, the offset is therefore exactly zero and $\eta_s=1$, however high the level.

\begin{table}[t]
\centering
\caption{Role of the ACF at the delay separation in the bound (Eq.~\eqref{eq:crlb_curv}), and in the offset (Eq.~\eqref{eq:bias_second}).}
\label{tab:roles}
\footnotesize
\setlength{\tabcolsep}{2.5pt}
\renewcommand{\arraystretch}{1.15}
\begin{tabular}{@{}>{\raggedright\arraybackslash}p{0.21\columnwidth}>{\raggedright\arraybackslash}p{0.13\columnwidth}>{\raggedright\arraybackslash}p{0.34\columnwidth}>{\raggedright\arraybackslash}p{0.26\columnwidth}@{}}
\toprule
ACF at $\Delta\tau$ & Couples & In the bound & In the offset \\
\midrule
Slope $\lambda_s$ & $\tau_\ell$, $\alpha_{\bar\ell}$ & removes delay information, $1-\eta_s$ & pulls the mainlobe \\
Curvature $\chi_s$ & $\tau_\ell$, $\tau_{\bar\ell}$ & couples the two delays, $\varrho_s$ & sharpens or flattens the mainlobe \\
Level $\bar g_s$ & $\alpha_\ell$, $\alpha_{\bar\ell}$ & amplifies the removed information, $1/(1-\bar g_s^{2})$ & raises or lowers the mainlobe \\
\bottomrule
\end{tabular}
\end{table}

\textit{Sidelobe Level and the Bound:} Fig.~\ref{fig:leakage_crlb_AB} shows $\sqrt{\mathrm{CRLB}(\Delta\tau)}$ for Group~A at $\mathrm{SNR}=10$~dB, with color indicating the normalized sidelobe level $\gamma_s=|\bar g_s|$, computed from the ACF of Eq.~\eqref{eq:gs}, i.e., from the allocation alone and not from the scan. For A3, the bound is lowest at $\Delta\tau = 2.17$~ ns, where $\gamma_s$ also reaches its first sidelobe maximum, $0.95$. At this maximum of the ACF, the slope vanishes and the curvature peaks together with the level. The delay information per bound is therefore fully retained, i.e., $\eta_s = 1$, while the coupling of the two delays, $\varrho_s \approx \cos\phi\,\bar g_s$, is strongest. Because it is negative here, it makes the errors of the two delays positively correlated, so that they partly cancel in the separation. The bound on each delay alone increases, but the cancellation outweighs this, and the separation bound falls below $v_1 + v_2$, the value it would have without the coupling. At the next maximum, where $\bar g_s > 0$, the coupling is positive and raises the separation bound.

\textit{Sidelobe Level and Peak Selection:} On its own, the level does not displace the peaks of the scan, but it changes their height, and with it the estimated gains. By Eq.~\eqref{eq:twopath_scan}, the change depends on the phase and the gain ratio of the two MPCs. The sidelobes of the second MPC can lower the mainlobe of the first and raise its nearest sidelobe above it, so that a peak search takes this sidelobe for the mainlobe. If the first MPC is found correctly, its response, including its sidelobes, can be cancelled before the second is estimated, but the same can occur for every further pair of MPCs. The level therefore needs to be analyzed together with the bound and the offset. Its magnitude $\gamma_s$ is the worst case of this multipath sidelobe interference for MPCs of equal gain magnitude, reached when their gain ratio is real.

\textit{Precision, Accuracy and Practical Use:} The bound of Eq.~\eqref{eq:crlb_curv} gives the precision that locally unbiased estimators of the two-path model can reach under AWGN. The accuracy of a practical estimator also depends on its bias and on the peaks it selects, so a larger bound indicates degraded precision rather than degraded accuracy. The single-path scan, common on commodity CSI, is biased by the offset of Section~\ref{subsec:twoPath}, which persists without noise. For the separations in Table~\ref{tab:twopath_peak_offsets}, this offset exceeds the square root of the bound by more than an order of magnitude at $\mathrm{SNR}=10$~dB, so the error of the scan is set by the offset rather than by its variance. Within the two-path model, the allocation enters the bound, the offset and the level only through $g_s$, while the gains set their size and sign. The slope, the curvature and the level are thus fixed by the allocation as functions of the separation, which is not known in advance. Before a measurement, they allow legal aggregations to be compared over the range of excess delays a deployment expects, by how much of this range is exposed to large offsets, lost delay information or competing sidelobes. After a measurement, the separations between the peaks found are known, and the level at these separations can indicate which peaks may be sidelobes of others.

\subsection{\texorpdfstring{Extension to More Than Two Paths}{Extension to More Than Two Paths}}
\label{subsec:Lpath}

With $L$ MPCs, each MPC still interacts with every other through the ACF at their delay separation. What changes is how these interactions combine, which differs for the bound, the deterministic offset and the sidelobe level.

\textit{CRLB:} With $L>2$ MPCs, the FIM covers $L$ delays and $2L$ gain components. Each of its entries is formed from the inner product of the derivative vectors $\mathbf{d}_i$ and $\mathbf{d}_j$ of two parameters, as in Eq.~\eqref{eq:crlb_fim_general}, and since each parameter belongs to one MPC, it involves at most two MPCs. The bound, however, follows from the inverse of the FIM and depends on all MPCs. From inverse FIM, with the gains and any delays not of interest removed by the Schur complement, the bound can be read on any delay, on the separation of any two MPCs, or jointly on the separations of all MPCs from a reference one, such as the first. In the joint bound, the entries that relate the first and the third MPC, for instance, do not contain the second, but its effect still reaches their separation through the inverse. Since the separations are unaffected by the timing offset between an unsynchronized transmitter and receiver (Section~\ref{sec:CRLB}), the joint bound would apply directly to bistatic links. Computing the bound as if some MPCs were absent would give a lower, optimistic value, since it leaves out the information lost. The exception is when those MPCs lie at delay separations at which the level, slope and curvature are negligible.

\textit{Deterministic Matched-Filter Offset:} With $L>2$ MPCs, the offset would still follow from the same Newton step, with all MPCs in the scan. To first order, the offset of each MPC is the sum of the displacements by all others, each through its relative gain and the slope of the ACF at its delay separation from that MPC. As shown in Section~\ref{subsec:twoPath}, even for two MPCs the first order does not suffice, and with more MPCs the second order contains products of the contributions of different MPCs, since the level of one MPC changes how far the slope of another displaces the peak. It therefore does not reduce to a sum of the offsets each MPC would cause alone, but needs to follow from the same Newton step, now with all MPCs included (Appendix~\ref{app:bias}).

\textit{Normalized Sidelobe level:} With $L>2$ MPCs, the multipath sidelobe interference at the delay of an MPC is the coherent sum of the sidelobes of all other MPCs there. Its worst case follows from the sum of their sidelobe levels, weighted by their gain ratios, but how close the interference comes to it depends on the gains, phases and delay separations of all MPCs, and is left for future work. Sidelobes of several MPCs can also add up to a false peak that no two MPCs produce alone, so peak selection is more exposed as $L$ grows. The cancellation would then proceed MPC by MPC, preceded by a joint estimation of the gains and delays of the MPCs found so far, accurate enough for their responses to be removed. Developing such an estimator is the necessary next step, since the level affects the estimated gains most and leads to the global errors of a wrong peak.

\section{Conclusion}
\label{sec:conclusion}

This paper studied multipath delay estimation on Wi-Fi-compliant non-contiguous allocations, combining the CRLB of the delay separation with the delay-domain response of the allocation.

A larger aperture lowers the bound at every separation, even when the spectrum is split across distant subbands. The gap adds oscillations on the scale of the inverse subband center spacing: at small separations the gapped allocation is penalized against a contiguous one of the same aperture, and at large separations the two converge. Expressed through the ACF, the bound depends on three quantities at the separation: the slope, which removes delay information to the unknown gains, the curvature, which couples the two delays, and the level, which amplifies the loss. The bound is lowest at the sidelobe maxima where the coupling is negative.

The single-path scan, common on commodity CSI, carries a deterministic offset, predicted in closed form to within $21$~ps. For the analyzed channels, it exceeds the square root of the bound by more than an order of magnitude at an SNR of $10$~dB, so the error of the scan is set by the offset rather than by its variance. A wider aperture sharpens the peak but does not necessarily reduce the offset, since the gap decides where each MPC falls on the sidelobes of the other.

The sidelobe level also decides whether the right peaks are selected. High sidelobes change the peak heights and the estimated gains, so a peak search can take a sidelobe for the mainlobe, and finding the remaining MPCs calls for cancelling the response of each MPC found, pair by pair. The normalized level, a worst case that depends on the allocation alone, flags the separations at which this can occur.

The bound, the offset and the sidelobe level are evaluated here at the true parameters, and thus establish what an allocation permits for a known channel state. The next step is to carry them over to measured CSI, where the gains and delays are first estimated and then refined, and to analyze the error of estimators that cancel the responses iteratively, including their peak selection and the global errors of a wrong peak, which the local bound and offset do not capture.

\appendices

\section{Fisher Information Matrix Derivations}
\label{app:fim}

{\footnotesize

\subsection{\texorpdfstring{Derivative Vectors and FIM Entries}{Derivative Vectors and FIM Entries}}
\label{app:fim_entries}

For scenario $s$ and the parameter vector $\boldsymbol{\theta}$ of Eq.~\eqref{eq:crlb_theta}, the $6\times6$ FIM of Eq.~\eqref{eq:crlb_fim_general} is
\begingroup
\fontsize{8pt}{9.6pt}\selectfont
\setlength{\jot}{1pt}
\begin{equation}
\mathbf{I}_s(\boldsymbol{\theta})
=
\left[
\begin{array}{cc|cccc}
I_{11} & I_{12} & I_{13} & I_{14} & I_{15} & I_{16}\\
I_{21} & I_{22} & I_{23} & I_{24} & I_{25} & I_{26}\\ \hline
I_{31} & I_{32} & I_{33} & I_{34} & I_{35} & I_{36}\\
I_{41} & I_{42} & I_{43} & I_{44} & I_{45} & I_{46}\\
I_{51} & I_{52} & I_{53} & I_{54} & I_{55} & I_{56}\\
I_{61} & I_{62} & I_{63} & I_{64} & I_{65} & I_{66}
\end{array}
\right]
=
\left[
\begin{array}{c|c}
\mathbf{I}_{\tau\tau} & \mathbf{I}_{\tau\alpha}\\ \hline
\mathbf{I}_{\alpha\tau} & \mathbf{I}_{\alpha\alpha}
\end{array}
\right],
\label{eq:FIM6x6_block}
\end{equation}
\endgroup
with $I_{ij}=[\mathbf{I}_s(\boldsymbol{\theta})]_{ij}$ and the index order
\begin{equation}
(1,2,3,4,5,6)\ \leftrightarrow\ (\tau_1,\tau_2,\alpha_{1\mathrm{Re}},\alpha_{1\mathrm{Im}},\alpha_{2\mathrm{Re}},\alpha_{2\mathrm{Im}}).
\label{eq:FIM_index_order}
\end{equation}
An MPC with unit gain at delay $\tau$ produces on the occupied subcarriers $\mathcal{K}_s$ the CFR
\begin{equation}
\mathbf{v}_s(\tau)=\big[a_s[n]\,e^{-j2\pi f[n]\tau}\big]_{n\in\mathcal{K}_s},
\label{eq:signature}
\end{equation}
referred to as its signature. The signatures of the two MPCs are the columns of $\mathbf{A}_s\mathbf{X}_s(\boldsymbol{\tau})$ in Eq.~\eqref{eq:crlb_mu}, so that $\boldsymbol{\mu}_s=\alpha_1\mathbf{v}_s(\tau_1)+\alpha_2\mathbf{v}_s(\tau_2)$, and the derivative vectors $\mathbf{d}_i=\partial\boldsymbol{\mu}_s/\partial\theta_i$ are
\begin{equation}
\begin{aligned}
\mathbf{d}_1&=\alpha_1\mathbf{v}_s'(\tau_1), & \mathbf{d}_3&=\mathbf{v}_s(\tau_1), & \mathbf{d}_4&=j\,\mathbf{v}_s(\tau_1),\\
\mathbf{d}_2&=\alpha_2\mathbf{v}_s'(\tau_2), & \mathbf{d}_5&=\mathbf{v}_s(\tau_2), & \mathbf{d}_6&=j\,\mathbf{v}_s(\tau_2),
\end{aligned}
\label{eq:derivative_vectors}
\end{equation}
where $\mathbf{v}_s'(\tau)=\partial\mathbf{v}_s(\tau)/\partial\tau=-j2\pi\big[a_s[n]f[n]e^{-j2\pi f[n]\tau}\big]_{n\in\mathcal{K}_s}$.

By Eq.~\eqref{eq:derivative_vectors}, each entry $I_{ij}=\frac{2}{\sigma_s^{2}}\Re\{\mathbf{d}_i^{H}\mathbf{d}_j\}$ involves an inner product of two signatures or of their delay derivatives. Each delay derivative contributes a factor $-j2\pi f[n]$ on subcarrier $n$, and two signatures at different delays leave the phase $e^{-j2\pi f[n]\Delta\tau}$, so every such inner product is a multiple of one of the sums
\begin{equation}
Q_p(u)=\sum_{n\in\mathcal{K}_s}|a_s[n]|^{2}f[n]^{p}e^{-j2\pi f[n]u},
\label{eq:moments}
\end{equation}
with $p\in\{0,1,2\}$ the number of delay derivatives in the product, and $u=\Delta\tau$ for vectors of different MPCs and $u=0$ for vectors of the same MPC. For different MPCs,
\begin{equation}
\begin{aligned}
\mathbf{v}_s^{H}(\tau_1)\,\mathbf{v}_s(\tau_2)&=Q_0(\Delta\tau),\\
\mathbf{v}_s^{H}(\tau_1)\,\mathbf{v}_s'(\tau_2)&=-j2\pi\,Q_1(\Delta\tau),\\
\mathbf{v}_s'^{H}(\tau_1)\,\mathbf{v}_s'(\tau_2)&=(2\pi)^{2}\,Q_2(\Delta\tau),
\end{aligned}
\label{eq:signature_products}
\end{equation}
and for the same MPC the same products give $Q_0(0)$, $-j2\pi Q_1(0)$ and $(2\pi)^{2}Q_2(0)$. Exchanging $\tau_1$ and $\tau_2$ replaces $\Delta\tau$ by $-\Delta\tau$, and $Q_p(-\Delta\tau)=Q_p^{*}(\Delta\tau)$. The delay entries are then
\begin{equation}
I_{\ell\ell}=\frac{8\pi^{2}}{\sigma_s^{2}}|\alpha_\ell|^{2}Q_2(0),\qquad
I_{12}=\frac{8\pi^{2}}{\sigma_s^{2}}\Re\{\alpha_1^{*}\alpha_2\,Q_2(\Delta\tau)\},
\label{eq:fim_delay}
\end{equation}
with $\ell\in\{1,2\}$. The gain entries are $I_{33}=I_{44}=I_{55}=I_{66}=\frac{2}{\sigma_s^{2}}Q_0(0)$, $I_{34}=I_{56}=0$ for any mask, and
\begin{equation}
I_{35}=I_{46}=\frac{2}{\sigma_s^{2}}\Re\{Q_0(\Delta\tau)\},\qquad I_{45}=-I_{36}=\frac{2}{\sigma_s^{2}}\Im\{Q_0(\Delta\tau)\}.
\label{eq:fim_gain}
\end{equation}
The delay--gain entries, paired as complex numbers, are
\begin{equation}
\begin{aligned}
I_{13}+jI_{14}&=-j\tfrac{4\pi}{\sigma_s^{2}}\,\alpha_1Q_1(0), & I_{15}+jI_{16}&=-j\tfrac{4\pi}{\sigma_s^{2}}\,\alpha_1Q_1^{*}(\Delta\tau),\\
I_{25}+jI_{26}&=-j\tfrac{4\pi}{\sigma_s^{2}}\,\alpha_2Q_1(0), & I_{23}+jI_{24}&=-j\tfrac{4\pi}{\sigma_s^{2}}\,\alpha_2Q_1(\Delta\tau).
\end{aligned}
\label{eq:fim_mixed}
\end{equation}
The four entries $I_{13}$, $I_{14}$, $I_{25}$ and $I_{26}$ pair a delay with the gain of its own MPC and are proportional to $Q_1(0)=\sum_{n\in\mathcal{K}_s}|a_s[n]|^{2}f[n]$, which the frequency centring of Section~\ref{subsec:observations} sets to zero (Appendix~\ref{app:centering}). Since $\mathbf{I}_s(\boldsymbol{\theta})$ is symmetric, the entries below the diagonal follow from $I_{ji}=I_{ij}$.

\subsection{\texorpdfstring{Frequency Centring}{Frequency Centring}}
\label{app:centering}

The frequency centring of Section~\ref{subsec:observations} places the origin of the frequency axis at the mask-weighted centroid, so that $Q_1(0)=0$ and the entries $I_{13}$, $I_{14}$, $I_{25}$ and $I_{26}$ of Eq.~\eqref{eq:fim_mixed} vanish. Each delay is then coupled only to the gain of the other MPC, and the allocation enters the FIM through $Q_0(0)$, $Q_2(0)$ and the three sums $Q_p(\Delta\tau)$ at the delay separation. The choice carries no information: displacing the origin by $f_0$ maps the gains as $\alpha_\ell\to\alpha_\ell e^{j2\pi f_0\tau_\ell}$, a map invertible at every $\boldsymbol{\tau}$ that leaves the delays untouched, so the effective information obtained from Eq.~\eqref{eq:crlb_fim_eff} is the same on either axis.

Centring constrains the first moment only and holds for any mask. A mask symmetric about the origin, as for the equal-bandwidth allocations considered in this paper, additionally makes $Q_0(u)$ and $Q_2(u)$ real and even and $Q_1(u)$ imaginary and odd, and sets every odd-order sum $\sum_{n\in\mathcal{K}_s}|a_s[n]|^{2}f[n]^{p}$ to zero. This symmetry is used in Appendix~\ref{app:close_sep}, in the reduction of $\psi_\ell$ in Section~\ref{subsec:twoPath}, and in Appendix~\ref{app:deriv}.

\subsection{\texorpdfstring{Effective Delay Information}{Effective Delay Information}}
\label{app:ieff}

The entries of $\mathbf{I}_{\mathrm{eff}}(\boldsymbol{\tau})$ in Eq.~\eqref{eq:crlb_fim_eff} are $\tilde{I}_{ij}=I_{ij}-[\mathbf{I}_{\tau\alpha}\mathbf{I}_{\alpha\alpha}^{-1}\mathbf{I}_{\alpha\tau}]_{ij}$, $i,j\in\{1,2\}$, the delay entries of the full FIM reduced by the part that the unknown gains absorb. Since $\boldsymbol{\mu}_s=\mathbf{A}_s\mathbf{X}_s(\boldsymbol{\tau})\boldsymbol{\alpha}$ is linear in the gains, the gain derivatives $\mathbf{d}_3,\dots,\mathbf{d}_6$ span the column space of $\mathbf{A}_s\mathbf{X}_s(\boldsymbol{\tau})$, and the Schur complement is the inner product of Eq.~\eqref{eq:crlb_fim_general} taken after projecting $\mathbf{d}_1$ and $\mathbf{d}_2$ off that space,
\begin{equation}
\begin{aligned}
\tilde{I}_{ij}&=\frac{2}{\sigma_s^{2}}\Re\big\{\mathbf{d}_i^{H}(\mathbf{I}_{N_s}-\mathbf{P}_s)\,\mathbf{d}_j\big\},\\
\mathbf{P}_s&=\mathbf{A}_s\mathbf{X}_s\big(\mathbf{X}_s^{H}\mathbf{A}_s^{H}\mathbf{A}_s\mathbf{X}_s\big)^{-1}\mathbf{X}_s^{H}\mathbf{A}_s^{H}.
\end{aligned}
\label{eq:ieff_proj}
\end{equation}
With $Q_1(0)=0$, the inner products in Eq.~\eqref{eq:signature_products} give
\begin{equation}
\begin{gathered}
\mathbf{X}_s^{H}\mathbf{A}_s^{H}\mathbf{A}_s\mathbf{X}_s=\begin{bmatrix}Q_0(0)&Q_0(\Delta\tau)\\ Q_0^{*}(\Delta\tau)&Q_0(0)\end{bmatrix},\\
\mathbf{X}_s^{H}\mathbf{A}_s^{H}\mathbf{d}_1=-j2\pi\alpha_1\begin{bmatrix}0\\ Q_1^{*}(\Delta\tau)\end{bmatrix},\qquad
\mathbf{X}_s^{H}\mathbf{A}_s^{H}\mathbf{d}_2=-j2\pi\alpha_2\begin{bmatrix}Q_1(\Delta\tau)\\ 0\end{bmatrix},
\end{gathered}
\label{eq:gram_proj}
\end{equation}
so that
\begin{equation}
\begin{aligned}
\tilde{I}_{\ell\ell}&=\frac{8\pi^{2}|\alpha_\ell|^{2}}{\sigma_s^{2}}\bigg[Q_2(0)-\frac{Q_0(0)\,|Q_1(\Delta\tau)|^{2}}{Q_0^{2}(0)-|Q_0(\Delta\tau)|^{2}}\bigg],\\
\tilde{I}_{12}&=\frac{8\pi^{2}}{\sigma_s^{2}}\Re\bigg\{\alpha_1^{*}\alpha_2\bigg[Q_2(\Delta\tau)+\frac{Q_0^{*}(\Delta\tau)\,Q_1^{2}(\Delta\tau)}{Q_0^{2}(0)-|Q_0(\Delta\tau)|^{2}}\bigg]\bigg\}.
\end{aligned}
\label{eq:ieff_entries}
\end{equation}
The denominator is the determinant of $\mathbf{X}_s^{H}\mathbf{A}_s^{H}\mathbf{A}_s\mathbf{X}_s$. The first terms in the brackets give $I_{\ell\ell}$ and $I_{12}$ of Eq.~\eqref{eq:fim_delay}. The second terms are the part that the unknown gains absorb. They are built from the correlation $Q_0(\Delta\tau)$ of the two signatures and from $Q_1(\Delta\tau)$, which measures how much a shift of one delay resembles a change in the gain of the other MPC, and they vanish where $Q_1(\Delta\tau)=0$.

\subsection{Close-Separation Limit of the Bound}
\label{app:close_sep}

This subsection derives Eq.~\eqref{eq:crlb_close}. For small $\Delta\tau$, the exponential in Eq.~\eqref{eq:moments} is expanded to fourth order in $2\pi\Delta\tau$, which brings in the same sums with $p=3$ and $p=4$ at $u=0$. For the symmetric mask of Appendix~\ref{app:centering}, $Q_1(0)=Q_3(0)=0$, and with $x=(2\pi\Delta\tau)^{2}$,
\begin{equation}
\begin{aligned}
Q_0(\Delta\tau)&=Q_0(0)-\tfrac{x}{2}\,Q_2(0)+\tfrac{x^{2}}{24}\,Q_4(0)+O(x^{3}),\\
Q_1(\Delta\tau)&=-j2\pi\Delta\tau\big[Q_2(0)-\tfrac{x}{6}\,Q_4(0)\big]+O(\Delta\tau^{5}),\\
Q_2(\Delta\tau)&=Q_2(0)-\tfrac{x}{2}\,Q_4(0)+O(x^{2}).
\end{aligned}
\label{eq:moment_expansion}
\end{equation}
Substituting into Eq.~\eqref{eq:ieff_entries}, the zeroth-order terms cancel in both brackets, which reduce to $\tfrac{x}{4}V_s$ and $-\tfrac{x}{4}V_s$, with $V_s=Q_4(0)-Q_2^{2}(0)/Q_0(0)$ the sum in Eq.~\eqref{eq:crlb_close}. To leading order,
\begin{equation}
\mathbf{I}_{\mathrm{eff}}(\boldsymbol{\tau})=\frac{8\pi^{4}\Delta\tau^{2}\,V_s}{\sigma_s^{2}}
\begin{bmatrix}|\alpha_1|^{2}&-\Re\{\alpha_1^{*}\alpha_2\}\\-\Re\{\alpha_1^{*}\alpha_2\}&|\alpha_2|^{2}\end{bmatrix}+O(\Delta\tau^{4}).
\label{eq:ieff_close}
\end{equation}
The matrix has determinant $|\alpha_1|^{2}|\alpha_2|^{2}-\Re\{\alpha_1^{*}\alpha_2\}^{2}=\Im\{\alpha_1^{*}\alpha_2\}^{2}$, and the numerator of Eq.~\eqref{eq:crlb_delta_tau} is proportional to $|\alpha_1|^{2}+|\alpha_2|^{2}-2\Re\{\alpha_1^{*}\alpha_2\}=|\alpha_2-\alpha_1|^{2}$. Eq.~\eqref{eq:crlb_delta_tau} therefore gives Eq.~\eqref{eq:crlb_close}, with
\begin{equation}
\kappa(\alpha_1,\alpha_2)=\frac{2\,|\alpha_2-\alpha_1|^{2}}{\Im\{\alpha_1^{*}\alpha_2\}^{2}}.
\label{eq:kappa}
\end{equation}
The factor $\kappa$ depends only on the path gains and is common to all allocations, so comparisons between allocations do not involve it. It diverges when the two gains are collinear, $\Im\{\alpha_1^{*}\alpha_2\}=0$, where the matrix in Eq.~\eqref{eq:ieff_close} is singular and the bound follows a higher-order law in $\Delta\tau$. The sum $V_s$ is nonnegative and vanishes when $f[n]^{2}$ is constant over the occupied subcarriers, which is approached by allocations whose subcarriers are grouped around two widely separated subband center frequencies.

\section{\texorpdfstring{Delay-Domain Offset and Curvature Relations}{Delay-Domain Offset and Curvature Relations}}
\label{app:deriv}

\subsection{\texorpdfstring{Deterministic Offset of the Single-Path Scan}{Deterministic Offset of the Single-Path Scan}}
\label{app:bias}

This appendix subsection derives Eqs.~\eqref{eq:bias_factored} and~\eqref{eq:bias_second}. Differentiating Eq.~\eqref{eq:gs} brings down a factor $j2\pi f[n]$ at each order:
\begin{equation}
\label{eq:gs_deriv}
g_s^{(p)}(u)=(j2\pi)^p\!\!\sum_{n\in\mathcal{K}_s}\!\!|a_s[n]|^{2}\,f[n]^{p}\,e^{j2\pi f[n]u},
\end{equation}
and for the symmetric mask of Appendix~\ref{app:centering}, $g_s$ is real and:
\begin{equation}
\label{eq:gs_curv0}
g_s'(0)=0, \qquad g_s''(0)=-(2\pi)^{2}\,g_s(0)\,B_{\mathrm{rms}}^{2},
\end{equation}
with $B_{\mathrm{rms}}$ as defined in Eq.~\eqref{eq:bias_defs}. With $\lambda_s$ of the same equation, this gives the identity $g_s'^{\,2}(u)=\lambda_s^{2}(u)\,g_s(0)\,|g_s''(0)|$.

The scan of Eq.~\eqref{eq:twopath_scan} is $T_s(\tau)=|F(\tau)|$, with $F(\tau)=\alpha_1 g_s(\tau-\tau_1)+\alpha_2 g_s(\tau-\tau_2)$, and its peaks are roots of:
\begin{equation}
\label{eq:stationarity}
\Theta(\tau)\;=\;\tfrac{1}{2}\,\frac{d}{d\tau}|F(\tau)|^{2}=\Re\{F^{*}(\tau)\,F'(\tau)\}.
\end{equation}
At $\tau_\ell$, where $\tau_\ell-\tau_{\bar\ell}=\mp\Delta\tau$ with the upper sign at $\tau_1$, $F$ and its derivatives are:
\begin{equation}
\label{eq:F_at_taul}
\begin{aligned}
F(\tau_\ell)&=\alpha_\ell\, g_s(0)+\alpha_{\bar\ell}\, g_s(\mp\Delta\tau),\\
F'(\tau_\ell)&=\alpha_{\bar\ell}\, g_s'(\mp\Delta\tau),\\
F''(\tau_\ell)&=\alpha_\ell\, g_s''(0)+\alpha_{\bar\ell}\, g_s''(\mp\Delta\tau),
\end{aligned}
\end{equation}
where the own-path term is absent from $F'(\tau_\ell)$ since $g_s'(0)=0$. Substituting Eq.~\eqref{eq:F_at_taul} into Eq.~\eqref{eq:stationarity} and into its derivative $\Theta'=|F'|^{2}+\Re\{F^{*}F''\}$, with $\arg(\alpha_{\bar\ell}/\alpha_\ell)=\pm\phi$ and the identity above, gives:
\begin{equation}
\label{eq:Phi_pair}
\begin{aligned}
\Theta(\tau_\ell)&=|\alpha_\ell|^{2}g_s(0)\,g_s'(\mp\Delta\tau)\,\bigl(r_\ell\cos\phi+r_\ell^{2}\,\bar g_s\bigr),\\
\Theta'(\tau_\ell)&=|\alpha_\ell|^{2}g_s(0)\,g_s''(0)\,\bigl[1+r_\ell\cos\phi\,(\chi_s+\bar g_s)\\
&\qquad+r_\ell^{2}\,(\bar g_s\chi_s-\lambda_s^{2})\bigr],
\end{aligned}
\end{equation}
with $r_\ell$, $\bar g_s$ and $\chi_s$ of Section~\ref{subsec:twoPath}. Both are quadratic in $r_\ell$: the $r_\ell^{2}$ term of $\Theta(\tau_\ell)$ is the other MPC acting on its own sidelobe, and that of $\Theta'(\tau_\ell)$ carries $\lambda_s^{2}$ from $|F'(\tau_\ell)|^{2}$. A single Newton step from $\tau_\ell$, $\hat b_\ell\approx-\Theta(\tau_\ell)/\Theta'(\tau_\ell)$, keeping only the leading term of each, gives the first-order offset:
\begin{equation}
\label{eq:bias_slope}
\hat b^{(1)}_\ell=-\frac{r_\ell\cos\phi\;g_s'(\mp\Delta\tau)}{g_s''(0)},
\end{equation}
which is Eq.~\eqref{eq:bias_factored} by Eqs.~\eqref{eq:bias_defs} and~\eqref{eq:gs_curv0}, since $\cos\phi\,g_s'(\mp\Delta\tau)=|g_s'(\Delta\tau)|\cos\psi_\ell$ for real $g_s$. Keeping all terms, the same Newton step gives the second-order offset:
\begin{equation}
\begin{aligned}
\hat b_\ell&=-\frac{g_s'(\mp\Delta\tau)}{g_s''(0)}\\
&\quad\times\frac{r_\ell\cos\phi+r_\ell^{2}\,\bar g_s}{1+r_\ell\cos\phi\,(\chi_s+\bar g_s)+r_\ell^{2}\,(\bar g_s\chi_s-\lambda_s^{2})},
\end{aligned}
\end{equation}
which, with Eq.~\eqref{eq:bias_slope} factored out, is Eq.~\eqref{eq:bias_second}.

With the gain of path $\ell$ known, the scan is replaced by $\Re\{\alpha_\ell^{*}F(\tau)\}$, whose derivative and curvature at $\tau_\ell$ are, by Eq.~\eqref{eq:F_at_taul}, $|\alpha_\ell|^{2}r_\ell\cos\phi\,g_s'(\mp\Delta\tau)$ and $|\alpha_\ell|^{2}g_s''(0)\,(1+r_\ell\cos\phi\,\chi_s)$. The same Newton step then gives $\hat b^{(1)}_\ell/(1+r_\ell\cos\phi\,\chi_s)$, whose first order is again Eq.~\eqref{eq:bias_factored} and, for the first MPC, the real part of~\cite[Eq.~(21)]{dun2021design}.

The remaining error of Eq.~\eqref{eq:bias_second} is that of linearizing $\Theta$ over the displacement. The Newton step also presumes a single root near $\tau_\ell$, which fails once the two mainlobes merge and no per-path offset is defined.

\subsection{\texorpdfstring{Closed Form of the Bound through ACF:}{Closed Form of the Bound trough ACF:}}
\label{app:ripple}

To express the bound through the ACF, the sums of Eq.~\eqref{eq:moments} can be compared with the derivatives in Eq.~\eqref{eq:gs_deriv}, which gives $Q_p(u)=g_s^{(p)}(-u)/(j2\pi)^{p}$. Hence $Q_0(0)=g_s(0)$ and, by Eq.~\eqref{eq:gs_curv0}, $(2\pi)^{2}Q_2(0)=|g_s''(0)|$. For the symmetric mask of Appendix~\ref{app:centering}, $Q_0(\Delta\tau)=\bar g_s\,Q_0(0)$ and $Q_2(\Delta\tau)=\chi_s\,Q_2(0)$ are real and $Q_1(\Delta\tau)$ is imaginary, with $Q_1^{2}(\Delta\tau)=-\lambda_s^{2}\,Q_0(0)\,Q_2(0)$ by the identity of Appendix~\ref{app:bias}. The second terms in the brackets of Eq.~\eqref{eq:ieff_entries} then become $\lambda_s^{2}Q_2(0)/(1-\bar g_s^{2})$ and $-\bar g_s\lambda_s^{2}Q_2(0)/(1-\bar g_s^{2})$. With $\Re\{\alpha_1^{*}\alpha_2\}=|\alpha_1||\alpha_2|\cos\phi$ and $c=2/\sigma_s^{2}$, for which $8\pi^{2}Q_2(0)/\sigma_s^{2}=c\,|g_s''(0)|$, the entries read:
\begin{equation}
\begin{aligned}
\tilde{I}_{\ell\ell}&=c\,|g_s''(0)|\,|\alpha_\ell|^{2}\Big(1-\frac{\lambda_s^{2}}{1-\bar g_s^{2}}\Big),\\
\tilde{I}_{12}&=c\,|g_s''(0)|\,|\alpha_1||\alpha_2|\cos\phi\Big(\chi_s-\frac{\bar g_s\,\lambda_s^{2}}{1-\bar g_s^{2}}\Big),
\end{aligned}
\label{eq:ieff_acf}
\end{equation}
so that $\tilde{I}_{\ell\ell}=c\,|g_s''(0)|\,|\alpha_\ell|^{2}\eta_s$ and $\tilde{I}_{12}/\sqrt{\tilde{I}_{11}\tilde{I}_{22}}=\varrho_s$, with $\eta_s$ and $\varrho_s$ of Eq.~\eqref{eq:rho_curv}. Inserting them into Eq.~\eqref{eq:crlb_individual} gives the individual bounds $v_\ell/[\eta_s(1-\varrho_s^{2})]$, with $v_\ell=1/(c\,|\alpha_\ell|^{2}|g_s''(0)|)$. Inserting them into Eq.~\eqref{eq:crlb_delta_tau} gives:
\begin{equation*}
\mathrm{CRLB}(\Delta\tau)=\frac{|\alpha_1|^{2}+|\alpha_2|^{2}+2|\alpha_1||\alpha_2|\,\varrho_s}{c\,|g_s''(0)|\,\eta_s\,|\alpha_1|^{2}|\alpha_2|^{2}(1-\varrho_s^{2})},
\end{equation*}
which, with $v_\ell$, is Eq.~\eqref{eq:crlb_curv}.

With the gains known, the second terms in the brackets of Eq.~\eqref{eq:ieff_entries} are absent, and wherever the slope vanishes, $Q_1(\Delta\tau)=0$, they are zero; in both cases $\eta_s=1$ and $\varrho_s=\cos\phi\,\chi_s$. For an asymmetric allocation the sums are complex, and the bound is evaluated from Eqs.~\eqref{eq:ieff_entries} and~\eqref{eq:crlb_delta_tau} directly.

\bibliographystyle{IEEEtran}
\bibliography{references.bib}

\end{document}